\documentclass[12pt]{article}
\advance\textheight by \topskip
\usepackage{graphicx}
\begin{document}
\thispagestyle{empty}
\begin{flushright}
{SU--ITP--93--29}\\
{IEM--FT--77/93}\\
astro-ph/9312039\\
16 December 1993 \\
\end{flushright}
\begin{center}
{\Large\bf Fluctuations  of the Gravitational Constant\\
\vskip .5cm
in the Inflationary Brans--Dicke Cosmology}\\
\vskip 1.1cm
{\bf Juan Garc\'{\i}a--Bellido}\footnote{
E-mail: bellido@slacvm.slac.stanford.edu},\\
\vskip .4cm
 {\bf Andrei Linde}\footnote{On leave  from:  Lebedev
Physical Institute, Moscow, Russia.\
E-mail: linde@physics.stanford.edu},
\vskip 0.05cm
Department of Physics, Stanford University, Stanford, CA 94305-4060,
USA
\vskip .5cm
{\bf Dmitri Linde}\footnote{E-mail: dmitri@cco.caltech.edu}
\vskip 0.05cm
{California Institute of Technology,
Pasadena, CA 91125, USA}
\end{center}
\vskip .8cm

{\centerline{\large\bf Abstract}}
According to the Brans--Dicke theory, the value of the
gravitational constant $G$ which we measure at present is
determined by the value of the Brans--Dicke scalar field $\phi$
at the end of inflation.  However, due to quantum fluctuations
of the scalar fields produced during inflation, the
gravitational constant $G(\phi)$ may take different values in
different exponentially large parts of the Universe.  We
investigate the probability distribution $P_p$ to find a domain
of a given volume with a given value of the gravitational
constant $G$ at a given time.  The investigation is performed
for a wide class of effective potentials of the scalar field
$\sigma$ which drives inflation, and with two different time
parametrizations.  Our work is based on the analytical study of
the diffusion equations for $P_p$, as well as on the computer
simulation of stochastic processes in the inflationary Universe.
We have found that in some inflationary models the probability
distribution $P_p$ rapidly approaches a stationary regime.  The
shape of the distribution depends, however, on the choice of the
time parametrization.  In some other models the distribution
$P_p$ is not stationary. An interpretation of our results and of
all ambiguities involved is outlined, and a possible role of
anthropic considerations in determination of the gravitational
constant is discussed.
\vfill
\newpage
\tableofcontents
\newpage

\section{\label{1} Introduction}
One of the most amazing properties of inflationary cosmology is
the process of self-reproduction of inflationary domains of the
Universe (for a review see \cite{MyBook}). This process exists
in many versions of inflationary universe scenario, including
old inflation \cite{b51}, new inflation \cite{b52,b62}, chaotic
inflation \cite{b19} and extended inflation \cite{EtExInf}.
Self-reproduction of inflationary domains implies that there is
no end of the evolution of the Universe. This process divides
the Universe into many exponentially large domains with all
types of symmetry breaking and with all types of
compactification of space-time compatible with inflation.

The best way to describe the global structure of the Universe in
this scenario is provided by the stochastic approach to
inflation.  The original version of this approach \cite{b60} was
based on the investigation of the  distribution of probability
$P_c(\phi,t)$ to find a given field $\phi$ at a given time at a
given point. This approach was not well suited for the
investigation of the process of self-reproduction of the
Universe.  A more adequate approach is based on investigation of
the distribution of probability $P_p(\phi,t)$ to find a given
field $\phi$ at a given time in a given physical volume
\cite{b19,b20,Nambu}.

The most detailed study of the distribution $P_p$ was performed
recently in \cite{LLM}, where it was shown that in many
inflationary models including the models with the effective
potentials $\phi^n$ and $e^{\alpha\phi}$  this probability
distribution rapidly approaches a stationary regime. This means
that if one takes a section of the Universe at a given time $t$
and calculates the relative fraction of domains of the Universe
with given properties (with given density, with given values of
various scalar fields, etc.), the result will not depend on the
time $t$, both during inflation and after it.

This result represents a major deviation of inflationary
cosmology from the standard Big Bang paradigm. A lot of work is
still needed to verify this result and to obtain its consistent
interpretation in the context of quantum cosmology. One should
also study how our methods work in the context of more
complicated models, including several different scalar fields
\cite{KL,LM,LLM}.

A natural idea would be to apply our methods to extended
inflation scenario \cite{b90}, which is a hybrid of the
Brans--Dicke theory and old inflation. In this scenario one has
two scalar fields: the Brans--Dicke field $\phi$ and the
inflaton field $\sigma$.  However, during extended inflation
only the former evolves in time, which reduces the problem to
the one we have studied already.  Moreover, the value of the
scalar field $\phi$ after inflation in this theory typically is
determined not by the stochastic dynamics during inflation, but
either by the position of a pole of a non-minimal kinetic term
of the field $\phi$ or by the effective potential of this field
which should be introduced  into the theory in order to make it
consistent with observational data \cite{Hyper}.

In this paper we will consider theories of another type, which
are hybrids of the Brans--Dicke theory with new
\cite{Liddle,Hybrid} or chaotic inflation \cite{ExtChaot}. In
these models, especially in the chaotic inflation model of ref.
\cite{ExtChaot}, simultaneous stochastic evolution of the fields
$\phi$ and $\sigma$ is very nontrivial. One of the most
interesting consequences of this evolution is the division of
the Universe after inflation into exponentially large regions
with different values of the gravitational constant in each of
them. Indeed, the gravitational constant $G$ in the Brans--Dicke
theory is a function of the field $\phi$,
\begin{equation}\label{GMP}
G = M_{\rm p}^{-2}(\phi) = {\omega\over 2\pi\phi^2} \  ,
\end{equation}
where $\omega$ is the Brans--Dicke parameter,  $\omega > 500$
\cite{VIK,PNB}. The value of the scalar field $\phi$ practically
does not change after inflation. Therefore, investigation of the
distribution $P_p(\phi,t)$  gives  us the fraction of the
physical volume of the Universe where the gravitational constant
$G$ has the effective value (\ref{GMP}).

The   stationary character of the probability distribution
$P_p(\sigma)$ in the theory of one scalar field $\sigma$ in the
standard general relativity theory is closely related to the
existence of the Planck boundary $\sigma_p$, where the potential
energy density $V(\sigma)$ becomes comparable with the Planck
density $M_{\rm p}^4$.  Typically the distribution
$P_p(\sigma,t)$ rapidly moves towards large $\sigma$, for the
reason that the volume of domains with large $V(\sigma)$ grows
very fast. The distribution $P_p(\sigma,t)$ becomes stabilized
as it approaches the Planck boundary, where, as it is argued in
\cite{LLM}, the process of self-reproduction of inflationary
domains is less efficient.

In the Brans--Dicke theory the situation is somewhat different.
The Planck boundary is not a point, but a line
$\phi_p(\sigma_p)$, where
\begin{equation}\label{Planck}
V(\sigma_p) = M_{\rm p}^4(\phi_p) = {4\pi^2\over\omega^2}
\phi_p^4 \ .
\end{equation}
Therefore, after the distribution $P_p(\sigma,\phi;t)$
approaches the Planck boundary, it can still move along this
boundary.  One may encounter three different possibilities:

1) The distribution $P_p(\sigma,\phi;t)$ never becomes
stationary for any values of $\phi$ and $\sigma$. Nevertheless
the properties of domains of inflationary universe filled by the
fields $\phi$ and $\sigma$ do not depend on the time when these
domains were formed. This is the most profound stationarity
which is present in all inflationary models where the process of
self-reproduction of inflationary domains is possible. In ref.
\cite{LLM} we called this property {\it microstationarity}, or
{\it local stationarity}.

2) The distribution $P_p(\sigma,\phi;t)$ normalized over all
possible values of $\phi$ and $\sigma$ gradually approaches a
stationary regime. In ref. \cite{LLM} we called this property
{\it macrostationarity}, or {\it global stationarity}.

3) The maximum of the distribution $P_p(\sigma,\phi;t)$ runs
away to infinitely large (or to infinitely small) values of
$\phi$ and $\sigma$. As a result, the relative fraction of
volume filled by any finite  $\phi$ and $\sigma$ as compared
with the {\it total} volume of the Universe decreases in time,
which means that the  global stationarity is absent. However,
one may wish to exclude from consideration domains with
infinitely large (or infinitely small) values of $\phi$
and $\sigma$  and concentrate on some finite part of space
($\phi,\sigma$) (for example, on those values of $\phi$ and
$\sigma$ which are consistent with the existence of life as we
know it). Then in some cases one may find out that the
probability distribution $P_p(\sigma,\phi;t)$ normalized over
this part of space ($\phi,\sigma$) (or the ratio
{$P_p(\sigma,\phi;t)\over P_p(\sigma_0,\phi_0,t)$) approaches a
stationary regime. In this case we will speak of a {\it runaway
stationarity}.

As we will see, all these possibilities may be realized in the
inflationary Brans--Dicke cosmology, depending on the choice of
the effective potential $V(\sigma)$.

There exists another important issue to be discussed in this
paper.  Even if the  distribution $P_p(\sigma,\phi;t)$ does not
depend on time, it may depend on the choice of possible time
parametrizations \cite{LLM}. As we will see, in the Brans--Dicke
theory this dependence can be quite strong.

In Section \ref{2} of this paper we describe chaotic inflation
in the Brans--Dicke theory at the classical level, following ref.
\cite{ExtChaot}. In Section \ref{3} we derive the stochastic
equations describing the evolution of quantum fluctuations of
the scalar fields $\phi$ and $\sigma$ during inflation, as well
as the origin of adiabatic energy density perturbations from
these fluctuations. In Section \ref{repr} we describe the
phenomenon of self-reproduction of the inflationary universe in
the presence of both scalar fields as well as the stochastic
approach to inflation in Brans--Dicke theory. We study the
boundary conditions for $P_p$ in Section \ref{4}. These
equations are extremely complicated, and it is not always
possible to solve them analytically.  Therefore we performed a
computer simulation of the stochastic evolution of the scalar
fields during inflation. We describe these numerical simulations
and their results in Section \ref{Computer}. In Section
\ref{Example} we analyze the case of a constant vacuum energy
density, where runaway stationarity appears. For general
increasing potentials $V(\sigma)$, the probability distribution
in the physical frame will rapidly move  to large values of
$\sigma$, constituting what we called runaway solutions. These
non--stationary solutions are described in Section \ref{run}.
Imposing   boundary conditions  at large $\sigma$ or introducing
steep potentials  $V(\sigma)$ will give a stationary distribution,
as described both numerically an analytically in Section
\ref{stat}. In Section \ref{6} we derive the stochastic
equations using a different time parametrization, where instead
of the usual time $t$ we choose the time $\tau \sim \ln a(t)$
\cite{b60,LLM}. The methods of computer simulations of stochastic
evolution in time $\tau$ are different from the methods which we
use in our simulations of the evolution in time $t$. We describe
these methods and their results in Section \ref{comtau}. We
compare the results obtained in different time parametrizations
and make an attempt of their interpretation in Section \ref{9}.

As we already mentioned, under certain conditions the
probability distribution $P_p$ is stationary, i.e.
time-independent. This apparently gives us a possibility to
calculate the position of the maximum of the distribution $P_p$
and thus predict the most probable value of the gravitational
constant in the Universe. However, this idea involves some
ambiguous speculations.  The readers will find them in  the
Appendix.

\section{\label{2} Inflationary Brans--Dicke Cosmology}
In this section we will describe the classical evolution of the
inflaton field with a generic chaotic potential, in the context
of the Jordan--Brans--Dicke theory of gravity,
\begin{equation}\label{S}
{\cal S}=\int d^4x \sqrt{-g} \left[{M_{\rm p}^2(\phi)\over16\pi}
R - {1\over2}(\partial\phi)^2 -{1\over2}(\partial\sigma)^2 -
V(\sigma)\right]\ .
\end{equation}
Here $M_{\rm p}^2(\phi) = {2\pi\over\omega}\phi^2$, \,
$\phi^2/8\omega$ is
the Brans--Dicke field, and $\sigma$ is
the inflaton field. We will consider several different
potentials, including $V(\sigma) \sim {m^2\over 2}\sigma^2$,\,
${\lambda\over 4}\sigma^4$,\, ${1\over 4\lambda}(m^2 - \lambda
\sigma^2)^2$ and  ${m^2\over 2}\sigma^2+{\lambda\over4}\sigma^4\,
\log{\sigma\over\sigma_0}$. The theory (\ref{S}) looks similar
to the extended inflation model \cite{b90}. The difference,
which will be very important for us, is that in the extended
inflation scenario inflation occurs at $\sigma = 0$, whereas in
our case inflation occurs during the slow rolling of the field
$\sigma$ \cite{ExtChaot}. The complete equations of motion in a
FRW Universe with scalar fields $\phi$ and $\sigma$ are 
\begin{equation}\begin{array}{c}\label{EQM}
{\displaystyle
\left(D^2 + \frac{1}{4\omega} R\right)\phi(t) = 0 }\ ,\\[3mm]
{\displaystyle
D^2 \sigma(t) = - \frac{\partial V}{\partial\sigma} }\ ,\\[3mm]
{\displaystyle
H^2 + \frac{\kappa}{a^2} = \frac{4\omega}{3\phi^2} \left({1\over2}
\dot{\phi}^2+{1\over2}\dot{\sigma}^2 + V(\sigma)\right) - 2 H
\frac{\dot{\phi}}{\phi} }\ ,
\end{array}\end{equation}
 where  $k = \pm 1, 0$; $D^2 \equiv \partial^2/\partial t^2 + 3 H
\partial/\partial t - k^2/a^2$, and $H = \dot a/a$.
During inflation we can consistently use the slow roll--over
approximations, $V''(\sigma),\ k^2/a^2 \ll H^2$, $\ \ddot{\phi}
\ll H\dot{\phi} \ll H^2 \phi$, $\ {1\over2}\dot{\sigma}^2 + {1\over2}
\dot{\phi}^2 \ll V(\sigma)$. 
The equations of motion (\ref{EQM}) then simplify to
\begin{equation}\begin{array}{c}\label{SEQ}
{\displaystyle
\dot{\phi} = \frac{\phi}{\omega} H\ , \hspace{5mm}\hspace{5mm}
\dot{\sigma} = - \frac{V'(\sigma)}{3H}\ , }\\[3mm]
{\displaystyle
R = - 12 H^2\ , \hspace{5mm}\hspace{5mm} H^2 =
\frac{4\omega}{3\phi^2}
V(\sigma) }\ .
\end{array}\end{equation}
We will be most interested in theories with
$V(\sigma)=\frac{\lambda}{2n} \sigma^{2n}$,
for which equations (\ref{SEQ}) reduce to
\begin{equation}\begin{array}{c}\label{EVO}
{\displaystyle
\dot{\phi} = \frac{2}{n} \left(\frac{n\lambda}
{6\omega}\right)^{1/2} \sigma^n }\ ,\\[3mm]
{\displaystyle
\dot{\sigma} = -\frac{\phi}{\sigma} \left(
\frac{n\lambda}{6\omega}\right)^{1/2} \sigma^n }\ .
\end{array}\end{equation}
It follows that in those theories $\ \phi$ and $\ \varphi
\equiv \sqrt{2\over n}\ \sigma $ move along a circle
of constant radius in the plane $(\varphi,\phi)$. We can
parametrize the classical trajectory by polar coordinates
$(\phi(t), \varphi(t)) = (r \sin \theta(t), r \cos \theta(t))$
with constant $r$, and angular velocity
\begin{equation}\label{DXT}
\dot{\theta}(t) = \left(\frac{\lambda}{3\omega}\right)^{1/2}
\left(\frac{n}{2}\right)^{\frac{n-1}{2}}
[r\cos\theta(t)]^{n-1}\ .
\end{equation}
For $n=1$ we find solutions with a constant angular velocity
\begin{equation}\begin{array}{rl}\label{CLA}
\phi(t)=&{\displaystyle
r \ \sin\left(\theta_o + \frac{m}{\sqrt{3\omega}}t
\right) } \ ,\\[3mm]
\sigma(t)=&{\displaystyle
\frac{r}{\sqrt{2}}\ \cos\left(\theta_o +
\frac{m}{\sqrt{3\omega}}t
\right) } \ ,\\[4mm]
a(t)=&{\displaystyle
a_o\ \left(\frac{\sin\left(\theta_o+{m\over\sqrt{3\omega}}t
\right)}{\sin\theta_o}\right)^\omega} \ .
\end{array}\end{equation}
Here $0 < \theta_o < \pi/2$. These solutions in the interval
$\theta_o
{\sqrt{3\omega}\over
m}< t \ll {\pi\sqrt{3\omega}\over 2m}$  correspond to the usual
power--law behavior $a(t) \sim t^\omega$.  For $n\geq 2$, the
angular velocity decreases with time and the classical solutions
are more complicated, but inflation is still power-law. These
solutions are actually attractors of the complete equations of
motion (\ref{EQM}), for all $n$ \cite{DGL}.

We will now discuss initial and final conditions for the
inflationary universe. In the chaotic inflation scenario, the
most natural initial conditions for inflation are set at the
Planck boundary, $V(\sigma_p) = M_{\rm p}^4(\phi_p)$, beyond
which a classical space--time has no meaning and the energy
gradient of the inhomogeneities produced during inflation
becomes greater than the potential energy density, thus
preventing inflation itself.  The initial conditions for
inflation are thus defined at the curve $\phi_p =
\left(\frac{\omega}{2\pi}\right)^{1/2} \left( {\lambda\over2n}
\right)^{1/4}\sigma_p^{n/2}$. On the other hand, inflation will
end when the kinetic energy density of the scalar fields becomes
comparable with the potential energy density,
${1\over2}\dot{\phi}^2 + {1\over2}\dot{\sigma}^2 \simeq
V(\sigma)$ or $m^2 = V''(\sigma) = H^2(\phi,\sigma)$. 
This condition corresponds to the end of inflation boundary
$\sigma_e = {n\over2\sqrt{3\pi}} M_{\rm p}(\phi_e) =
{n\over\sqrt{6\omega}} \phi_e $.

\

We will also consider an inflaton potential which leads to
spontaneous symmetry breaking, $V(\sigma) = {\lambda\over4}
\left(\sigma_o^2 - \sigma^2\right)^2$, where $\sigma_o\equiv
{m\over\sqrt\lambda}$. In this theory  equations (\ref{SEQ})
are expressed as
\begin{equation}\begin{array}{rl}\label{SEV}
\dot\phi=&{\displaystyle
\left(\frac{\lambda}{3\omega}\right)^{1/2}
|\sigma_o^2 - \sigma^2| }\ ,\\[3mm]
\dot\sigma=&{\displaystyle
\left(\frac{\lambda}{3\omega}\right)^{1/2} \phi\sigma
\ \frac{|\sigma_o^2 - \sigma^2|}{\sigma_o^2 - \sigma^2} }\ .
\end{array}\end{equation}
The fields move approximately along a circle centered at $\sigma =
\sigma_o$, in clockwise direction if the initial condition is to
the left of the minimum of the potential and in anticlockwise
direction if it is to the right. In fact, for $\sigma \gg
\sigma_o$ they will soon approach the asymptotic solution to
(\ref{EVO}) with $n=2$, while for $\sigma\ll\sigma_o$, we find
the solution
\begin{equation}\begin{array}{rl}\label{PSA}
\phi(t)=&{\displaystyle
\frac{m^2}{\sqrt{3\omega\lambda}} \ t }\ ,\\[4mm]
\sigma(t)=&{\displaystyle
\exp\left(\frac{m^2}{6\omega} t^2\right)}\ ,\\[3mm]
a(t)=&a_0 \ t^\omega\ .
\end{array}\end{equation}
Inflation will now occur in two disjoint sectors, either to the
left or to the right of the minimum. The Planck boundary for
initial conditions is again defined by $V(\sigma) = M_{\rm
p}^4(\phi)$, while the end of inflation is given by the
condition ${1\over2}\dot{\sigma}^2 \simeq V(\sigma)$.
These two boundaries are defined by the curves
\begin{equation}\begin{array}{rl}\label{CRV}
\phi_p=&{\displaystyle
\left(\frac{\omega\sqrt{\lambda}}{4\pi}\right)^{1/2}
|\sigma_o^2 - \sigma_p^2|^{1/2}}\ ,\\[3mm]
\phi_e=&{\displaystyle
\left(\frac{3\omega}{2}\right)^{1/2}
{|\sigma_o^2 - \sigma_e^2|\over\sigma_e} }\ .
\end{array}\end{equation}

In the absence of any potential for $\phi$, the Brans--Dicke
field remains almost constant after inflation, and therefore the
Planck mass today is approximately given by its value at the end
of inflation, $M_{\rm p}\simeq\sqrt{2\pi\over\omega}\ \phi_e\ $.
On the other hand, the total amount of inflation is approximately
${a_e\over a_p}\simeq\left({\phi_e\over\phi_p}\right)^\omega$.

The value of the Brans--Dicke parameter $\omega$ is bounded by
post-Newtonian experiments \cite{VIK} and primordial
nucleosynthesis   \cite{PNB} to be very large, $\omega >
500$, and therefore it is appropriate to use the approximation
$\omega \gg 1$ in the following analysis.

\section{\label{3} Stochastic Methods}
In this section we will describe the stochastic evolution of the
scalar fields during inflation and the generation of adiabatic
energy density fluctuations from quantum fluctuations of these
fields.

\subsection{Quantum Fluctuations of Scalar Fields}
The classical scalar fields $\phi$ and $\sigma$ in de Sitter
space are perturbed by their own quantum fluctuations, which are
stretched beyond the horizon and act on the quasi--homogeneous
background fields like a  stochastic force. In order to
calculate this effect we will coarse--grain over a horizon
distance and split the scalar fields into long--wavelength
classical background fields $\bar{\phi}(x)$ and
$\bar{\sigma}(x)$ plus short--wavelength quantum fluctuations
with physical momenta $k/a > H$,
\begin{equation}\begin{array}{rl}\label{CGF}
\phi(\vec{x},t)=&{\displaystyle
\bar{\phi}(\vec{x},t) + \int d^3k \ \theta(k -
\varepsilon a H)\left[a_k u_k(x) + a^\dagger_k u^\ast_k(x)
\right] }\ ,\\[3mm]
\sigma(\vec{x},t)=&{\displaystyle
\bar{\sigma}(\vec{x},t) + \int d^3k' \ \theta(k'-
\varepsilon a H)\left[b_{k'}v_{k'}(x)+ b^\dagger_{k'}
v^\ast_{k'}(x)\right] }\ ,
\end{array}\end{equation}
where $\varepsilon$ is an arbitrary parameter that shifts the
scale for coarse--graining \cite{b60}. The physical results
turn out to be independent of the choice of $\varepsilon$.
The quantum fluctuations are assumed to satisfy the following
commutation relations
\begin{equation}\begin{array}{c}\label{COM}
[a_k, a^\dagger_{k'}] = [b_k, b^\dagger_{k'}] =
\delta^3(\vec k - \vec k') \ , \  \ \ \  [a_k, b^\dagger_{k'}] = 0 \
{}.

\end{array}\end{equation}

The exact solutions to the scalar fields' equations (\ref{EQM})
in de Sitter space with $V(\sigma) = {1\over2} m^2 \sigma^2$
are given by \cite{BD}
\begin{equation}\begin{array}{rl}\label{UVK}
u_k(x)=&{\displaystyle
{e^{i\vec{k}\cdot\vec{x}}\over(2\pi)^{3/2}}\ {H\eta\over2}
\sqrt{\pi\eta}\ H_\mu^{(2)}(k \eta)\ , \hspace{5mm}\hspace{5mm}
\mu^2 = \frac{9}{4} + \frac{3}{\omega} \simeq \frac{9}{4} }\ ,\\[4mm]
v_{k'}(x)=&{\displaystyle  
{e^{i\vec{k'}\cdot\vec{x}}\over(2\pi)^{3/2}}\ {H\eta\over2}
\sqrt{\pi\eta}\ H_\nu^{(2)}(k'\eta)\ , \hspace{5mm}\hspace{5mm}
\nu^2 = \frac{9}{4} - \frac{m^2}{H^2} \simeq \frac{9}{4} }\ ,
\end{array}\end{equation}
where $\eta = - (a H)^{-1}$ is the conformal time, and
$H^{(2)}_{3/2}(x) = - \sqrt{2\over\pi x} \exp(ix)
\left(1+{i\over x}\right)$. The amplitude of the quantum
fluctuations of $\phi$ and $\sigma$ can be computed as
\begin{equation}\begin{array}{c}\label{STEP}
{\displaystyle    
\delta\phi = \left(4\pi k^3 |u_k|^2\right)^{1/2} =
{H\over2\pi} }\ ,\\[3mm]
{\displaystyle
\delta\sigma = \left(4\pi k'^3 |v_{k'}|^2\right)^{1/2} =
{H\over2\pi} }\ ,
\end{array}\end{equation}
which coincides with the Gibbons--Hawking temperature.

These quantum fluctuations then act as a stochastic force on the
classical background fields. One could write the evolution of
the coarse--grained fields in the form of Langevin equations
\begin{equation}\begin{array}{c}\label{LAN}
{\displaystyle
\frac{\partial\phi}{\partial t} = - \frac{M_{\rm p}^2(\phi)}
{2\pi} \frac{\partial H}{\partial \phi} +
\frac{ H^{3/2}}{2\pi} \zeta(t) }\ ,\\[3mm]
{\displaystyle
\frac{\partial\sigma}{\partial t} = - \frac{M_{\rm p}^2(\phi)}
{4\pi} \frac{\partial H}{\partial \sigma} +
\frac{  H^{3/2}}{2\pi} \xi(t) }\ ,
\end{array}\end{equation}
where $\zeta$ and $\xi$ behave like an effective white noise
generated by quantum fluctuations, $\left\langle\zeta(t)\
\zeta(t')\right\rangle \\ = \left\langle\xi(t)\ \xi(t')
\right\rangle = \delta(t-t')$ and $\ \left\langle\zeta(t)\ \xi(t')
\right\rangle = 0$, which leads to a Brownian motion of the
classical scalar fields $\phi$ and $\sigma$, with a typical
step (\ref{STEP}).

We can alternatively describe the stochastic process in terms of
the probability distribution $P_c(\sigma,\phi;t)$. This
distribution describes the probability to find the fields $\phi$
and $\sigma$ at a given time $t$ in a given point. Equivalently,
it describes the probability to find, at a given time $t$ in the
domain with a given {\it comoving} (i.e.  non-expanding) volume,
the fields with mean values $\phi$ and $\sigma$.  As it was
shown by Starobinsky \cite{b60}, this probability distribution
satisfies  the Fokker--Planck equation,
\begin{equation}\begin{array}{rl}\label{FPE}
{\displaystyle \frac{\partial P_c}{\partial t}}=&{\displaystyle
\frac{\partial}{\partial\sigma} \left(\frac{M_{\rm p}^2(\phi)}{4\pi}
\frac{\partial H}{\partial\sigma} P_c + \frac{  H^{3/2}}{8\pi^2}
\frac{\partial (H^{3/2} P_c)}{\partial\sigma} \right) }\\[3mm]
+&{\displaystyle \frac{\partial}{\partial\phi}
\left(\frac{M_{\rm p}^2(\phi)}{2\pi}\frac{\partial H}{\partial
\phi} P_c+\frac{  H^{3/2}}{8\pi^2} \frac{\partial (H^{3/2}P_c)}
{\partial \phi}  \right) \equiv  - \frac{\partial J_\sigma}
{\partial\sigma}  - \frac{\partial J_\phi}{\partial\phi} }\ ,
\end{array}\end{equation}
where we have chosen the Stratonovich version of stochastic
processes. This equation can be interpreted as the continuity
equation ${\displaystyle\ {\partial P_c\over\partial t}+\nabla
\cdot J = 0\ }$ associated with the conservation of probability.
The first terms of each current correspond to the classical
drift forces for the fields $\phi$ and $\sigma$ (\ref{EVO}),
while the second terms correspond to the quantum diffusion due
to short--wavelength fluctuations (\ref{STEP}).

One can then compute the field correlations during inflation
with the help of this probability distribution, assuming that
$H$ is approximately constant,
\begin{equation}\begin{array}{rl}\label{CRR}
{\displaystyle
\frac{\partial}{\partial t}
\left\langle\phi^p\sigma^q\right\rangle =}& {\displaystyle
\frac{  H^3}{8\pi^2} p(p-1)
\left\langle\phi^{p-2}\sigma^q\right\rangle + \frac{4p}{3H}
\left\langle\phi^{p-2}\sigma^q V(\sigma)\right\rangle } \\[3mm]
+&{\displaystyle
\frac{  H^3}{8\pi^2} q(q-1) \left\langle\phi^p
\sigma^{q-2}\right\rangle - \frac{q}{3H}
\left\langle\phi^p\sigma^{q-1} V'(\sigma)\right\rangle}\ .
\end{array}\end{equation}
For example, for $V(\sigma)={1\over2} m^2 \sigma^2$, we find
\begin{equation}\begin{array}{rl}\label{CSP}
\left\langle\phi^2\right\rangle=&{\displaystyle
\frac{  H^2 \omega}{8\pi^2} \left(e^{\frac{2Ht}
{\omega}} - 1\right) \simeq \frac{  H^3}{4\pi^2} t,
\hspace{5mm}\hspace{5mm} Ht \ll \omega },\\[3mm]
\left\langle\sigma^2\right\rangle=&{\displaystyle
\frac{3  H^4}{8\pi^2m^2}\left(1-e^{-\frac{2m^2t}
{3H}}\right) \simeq \frac{  H^3}{4\pi^2} t,
\hspace{5mm}\hspace{5mm} Ht \ll \frac{H^2}{m^2} },
\end{array}\end{equation}
Note that the dispersion of both fields due to Brownian motion
is identical for a relatively large time interval.

We will now describe the origin of energy density perturbations
in the Universe from quantum fluctuations of the scalar fields
during inflation, and postpone the study of the
self--reproduction of the inflationary universe for the next
section.

\subsection{Energy Density Perturbations}
In this subsection we will describe the generation of adiabatic
energy density perturbations, when both fields $\sigma$ and
$\phi$ are included. The amplitude of these perturbations has
been computed previously in \cite{EDP,DGL}. We will give here an
alternative derivation.

The energy density perturbations  could have originated during
inflation in our model (\ref{S}), as quantum fluctuations of
both scalar fields that first left the horizon during inflation
and later reentered during the radiation or matter dominated
eras. The amplitude of those perturbations can be computed in
the Einstein frame ($\tilde{g}_{\mu\nu}=\phi^2 g_{\mu\nu}$) by
using the equality
\cite{LYT}
\begin{equation}\label{WPR}
\left.
\left(1+{2\over3}{\tilde{\rho}\over\tilde{\rho}+\tilde{p}}\right)
{\delta\tilde{\rho}\over\tilde{\rho}}\right|_{\rm 2HC} = \left.
\left(1+{2\over3}{\tilde{\rho}\over\tilde{\rho}+\tilde{p}}\right)
{\delta\tilde{\rho}\over\tilde{\rho}}\right|_{\rm 1HC} \ ,
\end{equation}
where 1HC corresponds to the time when the perturbations first
left the horizon and 2HC to the time when they reentered. Since
during inflation $\tilde{\rho}+\tilde{p}\simeq0$, the amplitude
of reentering perturbations can be written as
\begin{equation}\label{DRH}
\left.{\delta\tilde{\rho}\over\tilde{\rho}}\right|_{\rm 2HC}
\simeq {2\eta\over3} \left.{\delta\tilde{\rho}\over\tilde{\rho}+
\tilde{p}}\right|_{\rm 1HC} \ ,
\end{equation}
where $\eta = 2/3\ (3/5)$ for perturbations reentering during
the radiation (matter) era. During inflation, the pressure and
energy density in our theory (\ref{S}) in the Jordan frame have
the expressions
\begin{equation}\begin{array}{c}\label{PPR}
{\displaystyle
\rho = {1\over2}\dot{\phi}^2 + {1\over2}\dot{\sigma}^2 +
V(\sigma)}\ ,\\[3mm] {\displaystyle
p = {1\over2}\dot{\phi}^2 + {1\over2}\dot{\sigma}^2 - V(\sigma)\ ,}
\end{array}\end{equation}
and thus $\rho+p=\dot{\phi}^2 + \dot{\sigma}^2$. The adiabatic
energy density perturbations follow from the quantum
fluctuations of the fields. In the Einstein frame,
$\sqrt{\tilde{g}}\tilde{\rho}=\sqrt{g}\rho$,
\begin{equation}\label{DPR}
{\delta\tilde{\rho}\over\tilde{\rho}} = {V'(\sigma)\over
V(\sigma)}\ \delta\sigma - 4{\delta{\phi}\over\phi}\ .
\end{equation}
Using the equations of motion (\ref{SEQ}) we find (for the cold
matter
dominated Universe)
\begin{equation}\label{ETH}
 \left.{\delta\tilde{\rho}\over\tilde{\rho}}\right|_{\rm 2HC}
\simeq - {6\over 5} H\ \left.\frac{\dot{\phi}\delta\phi +
\dot{\sigma}\delta\sigma}{\dot{\phi}^2 + \dot{\sigma}^2}
\right|_{\rm 1HC} = - \left.{3 H^2\over 5\pi}\ \frac{\dot{\phi}
+ \dot{\sigma}} {\dot{\phi}^2 + \dot{\sigma}^2}\right|_{N_\lambda}\ ,
\end{equation}
 where $N_\lambda$ stands for the number of e-folds before
the end of inflation associated with the horizon crossing of a
particular wavelength. For perturbations of the size of the
present horizon, we must compute the last expression for
$N_\lambda \sim 65\ $ \cite{MyBook}. For a general potential
$V(\sigma)$, we can write the density perturbations (\ref{ETH})
as
\begin{equation}\label{PER}
 \left.{\delta\rho\over\rho} \simeq {24\over5}
{H(\sigma,\phi)\over M_{\rm p}^2(\phi)}{V(\sigma)\over
V'(\sigma)} \left({ 1 + \dot\phi/\dot\sigma \over
1 + (\dot\phi/\dot\sigma)^2} \right)\right|_{N_\lambda}\ .
\end{equation}
Note that in the large $\omega$ limit, $\dot\phi\ll\dot\sigma$
during the last stages of inflation, which ensures the
approximate equivalence of the Einstein and Jordan frames.
We then recover the usual expression \cite{MyBook}
\begin{equation}\label{PRE}
\left.{\delta\rho\over\rho} \simeq {24\over5}
{H(\sigma,\phi)\over M_{\rm p}^2(\phi)}{V(\sigma)\over V'(\sigma)}
\right|_{N_\lambda}\ ,
\end{equation}
where $M_{\rm p}$ is now $\phi$--dependent. For theories with
potentials of the type $\lambda \sigma^{2n}$, it behaves like
\begin{equation}\label{REP}
\left.{\delta\rho\over\rho} \simeq
{6\omega\over 5n\pi} \left({2\omega\lambda\over3n}\right)^{1/2}
{\sigma^{n+1}\over\phi^3}\right|_{N=65}\ .
\end{equation}
In the case of the theory $\lambda\sigma^4$, the density
perturbation (\ref{REP}) takes the usual $\sqrt{\lambda}$
dependence. However, for the theory $m^2\sigma^2/2$, the
perturbation on the horizon scale is given by
\begin{equation}\label{RPE}
{\delta\rho\over\rho} \sim   {50 \,m\over M_{\rm p}(\phi_e)}\ .
\end{equation}
Therefore, we note that the larger is the Planck mass at the end
of inflation in a given region of the Universe, the smaller will
be the density perturbation in this region for this model. We
will return to the discussion of this result at the end of
the paper.

\section{\label{repr} Self--reproduction of the Inflationary
Universe}
Quantum perturbations produced during inflation are responsible
not only for galaxy formation, but also for the process of
self-reproduction of the whole inflationary universe. This is
the important effect which we are going to consider here in the
context of Brans--Dicke cosmology. We will begin with an
elementary description of this effect, and then return to its
description in the context of the stochastic approach to
inflation.

\subsection{Elementary Considerations}
An important property of the inflationary universe is that
processes separated by distances $l > H^{-1}$ proceed
independently of one another \cite{MyBook}.  In this sense any
inflationary domain of initial radius exceeding $H^{-1}$ can be
considered as a separate mini-universe, expanding independently
of what occurs outside it, as a consequence of the ``no-hair"
theorem for de Sitter space.

During a typical time $H^{-1}$ each such domain expands $e$
times, and its volume grows $e^3 \approx 20$ times. This means
that this domain becomes divided into 20 independent
inflationary domains. The values of the classical fields $\phi$
and $\sigma$ inside each of these domains can be obtained by
solving classical equations of motion for these fields. However,
in addition to it one should take into account quantum
fluctuations of these fields, which become ``frozen'' during the
time $H^{-1}$. They have a typical amplitude ${H\over 2\pi}$,
but they may have different signs in each of the new 20 domains.
Let us assume that this amplitude is much greater than the
classical shift of the values of these fields during the time
$H^{-1}$. In this case in 5 out of 20 domains the field $\phi$
jumps towards its smaller values, and the field $\sigma$ jumps
towards its greater values. The same happens during the next
time $H^{-1}$.

In the context of theories with $V(\sigma) \sim \lambda
\sigma^{2n}$, this leads to a continuous process of recreation
of inflationary domains. Using the classical equations of motion
(\ref{EVO}), one finds that the condition that quantum diffusion
is more important than the classical drift of the fields $\phi$
and $\sigma$ is satisfied for
\cite{ExtChaot,JGB}
\begin{equation}\label{BIF}
{3M_{\rm p}^4(\phi)\over4\omega}\left(1+{\phi^2\over\varphi^2}
\right) < V(\sigma) < M_{\rm p}^4(\phi)\ ,
\end{equation}
where the last term corresponds to the Planck boundary. From
this equation it follows that the inflationary universe with
most natural initial conditions (i.e. not far from the Planck
boundary) enters eternal regime of self-reproduction. During
this regime the Universe becomes filled with all possible values
of the fields $\phi$ and $\sigma$, independently of their
initial values in the region (\ref{BIF}). Consequently, the
parts of the Universe where inflation ends will consist of many
exponentially large domains in which the Planck mass $M_{\rm
p}(\phi)$ and the gravitational constant $G(\phi)$ may take all
possible values from 0 to $\infty$.

This scenario, which we called eternal inflation \cite{b19},
deviates strongly from the standard Big Bang theory.  For
example, according to the standard theory  a closed Universe
should eventually collapse and disappear.  In our scenario a
closed Universe which has at least one inflationary domain with
a field $\phi$ in the interval (\ref{BIF}) will never disappear
as a whole.

Here one should make some comments to avoid terminological
misunderstandings which sometimes appear in the literature.
Eternal inflation does not mean that each part of the Universe
eternally inflates. The typical length of each geodesic at the
stage of inflation is finite. However, there is {\it no upper
limit} to the length of these geodesics, and those extremely
rare geodesics which have large length give the dominant (and
permanently growing) contribution to the total volume of the
Universe.  Therefore in our scenario inflation in the whole
Universe has no end, even though it ends on each particular
geodesic within a finite time.

One may try to reverse the question and ask whether inflationary
universe has any beginning. Unfortunately, the answer to this
question is much less definite. One may argue that each
geodesic being continued to the past has finite length (it
begins with a singularity) \cite{Nuffield}.  However, this is
not enough to prove that the Universe has   a single beginning
at some moment $t= 0$ in the past  (Big Bang). Whereas such a
possibility is not excluded, in order to prove it one should
show that there is an {\it upper limit} to the length of   all
geodesics continued to the past.  Indeed, even if   long
geodesics are extremely rare, they may give exponentially
large contribution to the present volume of the Universe.
At present we do not have any proof that that there exists any
upper limit to the length of  all geodesics continued to the past.
Therefore finiteness of length of each geodesic being 
continued to the past  \cite{Nuffield} does 
not mean yet that inflation is eternal only in future. We 
emphasize again that the length of each particular 
geodesic at the stage of inflation is also finite, and still we 
are speaking about eternal inflation.

On the other hand,  the properties of each particular
inflationary domain created in the process of self-reproduction
of the Universe do not depend on the   time when it was created
(microstationarity). Therefore by local observations which we
can make inside our domain  we cannot come to any conclusion
about the time when the Big Bang happened. Therefore, our
scenario removes the Big Bang to the indefinite past and in this
sense makes its possible existence almost irrelevant \cite{LLM}.
In particular, the stationary probability distributions which we
are going to obtain will not depend on initial conditions at the
beginning of our computer simulations. 

\subsection{Stochastic Approach}

A formal method to describe the process of  self-reproduction of
inflationary domains is given by the stochastic approach to
inflation. One of the possibilities is to go along the lines of
ref.  \cite{b20}, to solve the Fokker-Planck equation
(\ref{FPE}) for the distribution $P_c$, and then to study $P_p$
using these solutions.

For initial conditions of the fields $\sigma$ and $\phi$ far
away from the Planck boundary,  the probability distribution in
the comoving frame behaves like a Gaussian centered around the
classical trajectory $(\sigma_c(t),\phi_c(t))$ in the
$(\sigma,\phi)$ plane,
\begin{equation}\label{PRB}
P_c(\sigma,\phi;t) \sim \exp\left\{-\frac{(\sigma-\sigma_c(t))^2}
{2\Delta^2_\sigma(t)}-\frac{(\phi-\phi_c(t))^2}{2\Delta_\phi^2(t)}
\right\} \ ,
\end{equation}
with dispersion coefficients \cite{JGB}
\begin{equation}\begin{array}{rl}\label{DWZ}
\Delta^2_\sigma(t) =&{\displaystyle
\frac{\lambda}{3n^2} \left({\omega\over2\pi}\right)^2
\frac{\sigma_c^{2n-2}}{\phi_c^4} (\sigma_o^4-\sigma_c^4)
\sim \frac{\lambda}{3n^2} \left({\omega\over2\pi}\right)^2
\frac{\sigma_c^{2n-2} \sigma_o^4}{\phi_c^4} } \ ,\\[3mm]
\Delta^2_\phi(t) =&{\displaystyle
\frac{\lambda}{3n} \left({\omega\over2\pi}\right)^2
\sigma_c^{2n}\left(\frac{1}{\phi_o^2} - \frac{1}{\phi_c^2}\right)
\sim \frac{\lambda}{3n} \left({\omega\over2\pi}\right)^2
\frac{\sigma_c^{2n}}{\phi_o^2} \ .}
\end{array}\end{equation}

These results are very similar to the results of the
investigation of $P_c$ in the theory of a single scalar field
obtained in \cite{b20}. One may then use the fact that
during small time intervals $\Delta t$ the probability
distribution $P_p(\sigma,\phi;\Delta t)$, which takes into
account the difference of the rates of the quasi--exponential
growth of the proper volume in different parts of the domain, is
related to $P_c$ in a rather simple way,
\begin{equation}\label{PPC}
P_p(\sigma,\phi;\Delta t)\simeq P_c(\sigma,\phi;\Delta t) \
e^{3H\Delta t}\ .
\end{equation}

With the help of this approximate relation  one can study the
qualitative features of the behavior of $P_p$ and confirm the
existence of the regime of self-reproduction \cite{b20}.
However, to obtain a more detailed information about $P_p$ one
should study directly the  diffusion equation for $P_p$. This
equation differs from the equation for $P_c$ only by the
presence of an extra term $3HP_p$ \cite{Nambu,LLM}:
\begin{equation}\begin{array}{rl}\label{3HP}
{\displaystyle
\frac{\partial P_p}{\partial t} }=&{\displaystyle
\frac{\partial}{\partial\sigma}\left(\frac{M_{\rm p}^2(\phi)}
{4\pi} \frac{\partial H}{\partial\sigma} P_p + \frac{H^{3/2}}
{8\pi^2} \frac{\partial}{\partial\sigma}\left(H^{3/2}P_p \right)
\right)  } \\[3mm]
+&{\displaystyle
\frac{\partial}{\partial\phi}\left(\frac{M_{\rm p}^2(\phi)}
{2\pi} \frac{\partial H}{\partial\phi} P_p + \frac{H^{3/2}}
{8\pi^2} \frac{\partial}{\partial\phi}\left(H^{3/2}P_p \right)
\right) + 3HP_p \ . }
\end{array}\end{equation}

Apart from studying the distribution of fields $\sigma$ and
$\phi$ in all domains during inflation, we will calculate the
volume of all domains where inflation ends in a state with given
$\phi$ within each new time interval. This gives us  the
fraction of the volume of the Universe where inflation ends at a
given time $t$ within a given interval of values of the field
$\phi$. We call this new distribution ${\cal P}_p(\phi_e,t)$. This
distribution
is closely related to $P_p$. For
example, in the theories with $V(\sigma)=\frac{\lambda}{2n}
\sigma^{2n}$
\begin{equation}\label{PFP}
{\cal P}_p(\phi_e,t) \sim \phi_e^n \cdot P_p(\phi_e,\sigma_e,t) \ .
\end{equation}
Indeed, during the time $\Delta t$ all domains in the interval
$\Delta \sigma$ from  $\sigma_e - \dot{\sigma}\Delta t$ to
$\sigma_e $ will cross the boundary of the end of inflation at
$\sigma = \sigma_e$.  According to eq.  (\ref{EVO}), in the
theories we consider $\Delta \sigma = -\dot{\sigma}\, \Delta t=
\frac{\phi}{\sigma} \left(\frac{n\lambda}{6\omega}\right)^{1/2}
\sigma^n \, \Delta t$. The value of the field $\phi$ near the
end of inflation almost does not change, $\phi= \phi_e$, and   $
\sigma = \sigma_e = {n\over\sqrt{6\omega}} \phi_e$. This yields
$\Delta \sigma = \sqrt{\lambda\over n}{n^n\over
(6\omega)^{n/2}}\, \phi_e^n \Delta t$.  Obviously, the fraction
of the volume of the Universe where inflation ends at a given
time $t$ within a given interval of values of the field $\phi=
\phi_e$ is proportional to $P_p(\phi_e,\sigma_e,t)
\Delta\sigma$. This gives eq. (\ref{PFP}), up to an overall
normalization factor.

Since the value of the effective Planck mass $M_{\rm p}(\phi)$
after inflation practically does not change, this distribution
is most directly related to the fraction of the volume of the
post-inflationary universe with the Planck mass $M_{\rm p}
(\phi_e) \sim \sqrt{2\pi\over\omega}\ \phi_e$.
In this paper we will be interested mainly in  stationary distributions.
Whenever our distributions will be time-independent, we will write them simply
as $P_p(\phi)$ or
${\cal P}_p(\phi_e)$.

\section{\label{4} Stationary Probability Distributions}

In general it is very difficult to find any analytic
solutions to the equation (\ref{3HP}) for $P_p(\sigma,\phi;t)$.
However, in certain cases the corresponding solutions in the
limit of large $t$ can be represented in the simple form,
\begin{equation}\label{SPT}
P_p(\sigma,\phi;t) \sim e^{Et}\  \tilde P_p(\sigma,\phi)\ ,
\end{equation}
where $E$ is some constant \cite{Nambu,LLM}. In such cases the
normalized probability distribution $\tilde P_p(\sigma,\phi)$
will be stationary.\footnote{Since the difference between $P_p$
and  $\tilde P_p$ is only in the normalization, we will usually
omit the tilde and write $\tilde P_p$ simply as $P_p$.}
Analytical investigation of  $\tilde P_p(\sigma,\phi)$
often can be simplified if one studies instead the function
$\Psi(\sigma,\phi)$, where
\begin{equation}\label{PSI}
\tilde P_p(\sigma,\phi) \propto  H^{-3/2}(\sigma,\phi)
\exp\left(\frac{3M_{\rm p}^4(\phi)}{16V(\sigma)}
\right)\Psi(\sigma,\phi) \ .
\end{equation}
Using the  identity $D^2g - 2D(Df\ g) = e^f D^2(e^{-f} g) -
[(Df)^2 + D^2f]g$, one can show that the new function
$\Psi(\sigma,\phi)$ satisfies a two--dimensional
Sch\"odinger--like equation
\begin{equation}\label{FEP}
\left(H^{3/2}\frac{\partial}{\partial\sigma}\right)^2\Psi +
\left(H^{3/2}\frac{\partial}{\partial\phi}\right)^2\Psi -
V(\sigma,\phi) \Psi = 8\pi^2 E \Psi \
\end{equation}
with a new effective potential
\begin{equation}\begin{array}{rl}\label{VXY}
V(\sigma,\phi) =&{\displaystyle
\frac{16\pi^4}{9} H^{-5}(\sigma,\phi) V'(\sigma)^2 +
\frac{4\pi^2}{3} H^{-1}(\sigma,\phi) \left(\frac{5}{4}
\frac{V'(\sigma)^2}{V(\sigma)} - V''(\sigma)\right) }\\[3mm]
+&{\displaystyle \frac{64\pi^4}{3\omega} H^{-3}(\sigma,\phi)
V(\sigma) + \frac{6\pi^2}{\omega} H(\sigma,\phi) -
24\pi^2 H(\sigma,\phi) } \ .
\end{array}\end{equation}

This equation (or equation (\ref{3HP})) should be supplemented
with boundary conditions. There are three possible boundaries
in the $(\sigma,\phi)$ plane.

1. {\bf End of inflation boundary.}   Our diffusion
equations are valid only during inflation. Therefore some
boundary conditions should be imposed at the boundary where
inflation ends. These conditions follow from the continuity of
the probability distribution $P_c$ and of the probability
current $\vec J$ \cite{LLM}. In the theories with $V(\sigma)
\sim \sigma^{2n}$ and $\omega \gg 1$ the field $\phi$ at the end
of inflation almost does not change. The continuity condition
can be expressed  in terms of the field $\sigma$ changing from
the right side of the boundary $\sigma_e$ (from $\sigma_{e^+}$)
to the left of it (to $\sigma_{e^-}$):
\begin{equation}\label{CNT}
P_c(\sigma_{e^+}) = P_c(\sigma_{e^-})\ , \hspace{5mm}\hspace{5mm} \
\vec{J}(\sigma_{e^+}) = \vec{J}(\sigma_{e^-})\ .
\end{equation}
As it is shown in \cite{LLM}, this leads to the following
boundary condition on $P_p$:
\begin{equation}\label{VDB}
 \frac{\partial}{\partial\sigma}
\left(H^{3/2}(\sigma,\phi)\ P_p(\sigma,\phi)\right)|_{\rm end} = 0\ .
\end{equation}
One can re-express this boundary condition in terms of the
redefined function $\Psi(\sigma,\phi)$ (\ref{PSI}) as
\begin{equation}\label{BCC}
\left . \frac{\partial}{\partial\sigma} \left(e^{\frac{3M_{\rm
p}^4(\phi)}
{16 V(\sigma)}} \Psi(\sigma,\phi)\right)\right|_{\rm end} = 0\ .
\end{equation}

2. {\bf Planck boundary.}   The distribution
$P_p(\sigma,\phi;t)$ typically tends to be shifted towards the
region of greatest possible Hubble constant, which ensures
exponentially fast growth of volume of inflationary domains
$\sim e^{3H(\sigma,\phi)t}$.  However, one may argue that
inflation destroys itself   at values of the potential energy
density above the Planck scale by production of large gradients
of density.  Furthermore, the classical space--time in which
inflation takes place cease to make sense above the Planck
scale, where quantum fluctuations of the metric are important.
Therefore it is  natural to impose some boundary conditions at
the Planck boundary which would not allow a nonvanishing $P_p$
at densities higher than $M^4_p(\phi)$. As it is argued in
\cite{LLM}, most of the results are not very sensitive to a
particular choice of such boundary conditions (absorbing,
reflecting, etc.). Therefore we will simply assume that the
probability distribution $P_p(\sigma,\phi;t)$ vanishes when
$V(\sigma) = M_{\rm p}^4(\phi)$:
\begin{equation}\label{PP0}
P_p(\sigma_p,\phi_p;t) \propto \Psi(\sigma_p,\phi_p) = 0\ .
\end{equation}
Here $\sigma_p,\phi_p$ is any pair of values of the fields $\sigma$
and
$\phi$
belonging to the line $V(\sigma) = M_{\rm p}^4(\phi)$ (Planck
boundary).

3. {\bf Boundary at large $\sigma$.}   The two boundary
conditions mentioned above are not enough to ensure stationarity
of solutions, since the maximum of the probability distribution
may move along the Planck boundary. The reason is very simple.
In the ordinary inflationary theory the maximum of the
probability distribution moves towards the Planck boundary since
near this boundary the rate of exponential expansion of the
Universe is maximal. In our case the Planck boundary is not a
point where $V(\sigma) = M^4_p$ but a line $\phi(\sigma)$
(\ref{Planck}).  The greater is $\phi$ along this line, the
greater is the energy density there, the greater is the rate of
expansion. Therefore one may expect the probability distribution
$P_p$ to move along the Planck boundary towards greater and
greater values of $\phi$ and $\sigma$.

This would mean that there is no macrostationarity (global
stationarity) in our model, whereas the microstationarity (local
stationarity) is still present. Even though greater and greater
numbers of inflationary domains will contain indefinitely large
values of the fields, there will be exponentially many domains
with smaller values of these fields as well, and the properties
of these domains will not depend on the time $t$ when they are
created; see \cite{LLM} where this situation is discussed. In
such models we come to a peculiar conclusion that the main
fraction of the physical volume of the Universe is in a state
with an indefinitely large $M_{\rm p}$. This might  not be a
real problem, since life of our type simply cannot exist in the
parts of the Universe with too large  (and  too small) $M_{\rm
p}$.

Still it may be important to have global stationarity, see Appendix.
The simplest way to achieve it would be to impose
absorbing or reflecting boundary conditions at sufficiently
large $\sigma$, which would preclude the motion of the
distribution $P_p$ towards large $\sigma$. Such boundary
conditions are not unreasonable. Indeed, it is hard to expect
that in realistic theories inflation will be possible at all
indefinitely large values of $\phi$ and $\sigma$. It may happen,
for example, that the effective potential $V(\sigma)$ becomes
  steeper at large $\sigma$, and inflation (or at least the
process of self-reproduction of inflationary domains) becomes
impossible there. For example, one may consider a potential $V
\sim \lambda\sigma^n e^{\alpha^2\sigma^2}$. In this theory
inflation becomes impossible at $\sigma > \sigma_b \equiv
\alpha^{-1}$. As a result, the distribution $P_p$ acquires a
maximum somewhere near to the Planck boundary close to  $\sigma
= \sigma_b$. Another possibility which leads to a similar effect
is that the effective potential $V(\sigma)$ decreases at
sufficiently large $\sigma$, for example, $V \sim \lambda
\sigma^n e^{-\alpha^2\sigma^2}$. In this case the distribution
$P_p$ moves to large $\sigma$ until it reaches the maximum of
$V(\sigma)$.

As we will see from the results of our computer simulations, in
both cases the effect of the modification of $V(\sigma)$ at
large $\sigma$ can be mimicked by the introduction of a boundary
at some value of the field $\sigma = \sigma_b$. In our
analytical investigation of $P_p$ we will assume that
\begin{equation}\label{BCB}
\Psi(\sigma,\phi)|_{\sigma_{\phantom {}_b} }= 0\ .
\end{equation}
The way we impose boundary conditions in our numerical
investigations will be explained in the next  section.

\section{\label{Computer}Computer Simulations}
The stochastic equations (\ref{FEP}) are rather complicated
  partial differential equations, and it is not always
possible to obtain their solution analytically even in the
theories with one scalar field \cite{LLM}. In the two-field case
the situation is even more complicated. Therefore, instead of
solving these equations directly, we will make a computer
simulation of the processes we are trying to investigate.

The main idea of our simulations is the following. We consider
$N$ points in the $(\sigma,\phi)$ plane. Each such point
represents the value of the scalar field in a  region of size
$O(H^{-1}(\sigma,\phi))$ ($h$-region). Our calculations should
give us the function $P_p(\sigma,\phi)$, which is interpreted as
the number of $h$-regions with field values $\sigma$ and $\phi$.
In our figures this function looks as a two-dimensional surface
in a three-dimensional space $(\sigma,\phi,P_p)$.

The values of the fields in each point are initially set to
$(\sigma_i,\phi_i)$. Then we calculate the values of the fields
in each point independently, since each such point represents an
$h$-region causally disconnected from other $h$-regions
(`no-hair' theorem for de Sitter space).

Each step of our calculation corresponds to a time change
$\Delta t = u H^{-1}_0$, where $H_0 = H(\sigma_i,\phi_i)$, and
$u$ is some number, $u < 1$. (The results should not depend on
$u$ if it is small enough.)

The evolution of the fields in each domain consists of  several
independent parts.  First of all, each field evolves according
to classical equations of motion during inflation. Secondly,
each field makes quantum jumps by $\delta\sigma ={H  \over \pi}
\sqrt{u H \over 2H_0}\,  \sin r_1 $, $ \delta\phi =  {H  \over
\pi}  \sqrt{u H\over 2H_0}\, \sin r_2$. Here $r_1$ and $r_2$ are
random numbers which are different for each point.

To make a computer simulation of this branching process, we
follow each domain until it grows in size two times, and after
that we considered it as 8 independent
$h$-regions.\footnote{Note that this does not necessarily
correspond to taking steps  $\Delta t = H_0^{-1} \log 2$.
Indeed, in the domains with $H( \sigma,\phi) \gg H_0$ the size
of the domains grows two times during a time interval much
smaller than $H_0^{-1} \log 2$. } If  we would continue doing so
for a long time, the number of such regions (and our
distribution $P_p(\sigma,\phi)$) would grow exponentially, 
and it would be extremely difficult to continue the calculation. 
However, in order to obtain a correct probability distribution
 it is not necessary to follow all   domains, since all of them 
evolve absolutely independently. 
In order to obtain  a normalized distribution $P_p(\sigma,\phi)$,
after each step of the calculations we were randomly 
removing some of the domains, but we were doing it is such 
a way that the probability for any domain to be removed was 
proportional to the distribution $P_p(\sigma,\phi)$ at this step 
of calculation. This allowed us  to keep the total number of
domains fixed and the distribution $P_p(\sigma,\phi)$ properly 
normalized without changing at any stage of calculation 
the correct shape
of the distribution $P_p(\sigma,\phi)$.\footnote{There is some
subtlety here. The volume corresponding to each $h$-region is
proportional to $H^{-3}(\sigma,\phi)$. Thus, if we are
interested in the relative fraction of the {\it volume} of the
Universe, we should show in our figures not the total number of
$h$-regions with given values of the fields, but the total
number of such regions multiplied by $H^{-3}(\sigma,\phi)$.
However, typically the difference between these two
distributions is not important, since $P_p(\sigma,\phi)$ depends
on $\sigma$ and $\phi$ much stronger than
$H^{-3}(\sigma,\phi)$.}

A special care should be taken about the points near the
boundaries.  As we already mentioned in the previous subsection,
there are boundaries of three different types in our problem.

1) {\bf The end of inflation boundary.}   In the theories with
$V(\sigma) \sim \lambda \sigma^n$   this boundary is given by
$\sigma \sim {n \over\sqrt{6\omega}}\, \phi$. When the field
$\sigma$ inside a given $h$-region becomes smaller than ${n
\over \sqrt{6\omega}}\, \phi$, inflation in this domain ends,
and the value of the field $\phi$ (and of the gravitational
constant) in this domain practically does not change after that
moment. We discard all domains where this happens. Then we  add new ones
in order to preserve correct normalization of our distribution.  However, each
time when we are adding new domains, we distribute them with the probability
distribution
proportional to $P_p$ at that time.  As we already mentioned, 
this method allows us to keep
the distribution $P_p$ normalized at all times without  
distorting its shape.\footnote{To avoid misunderstandings, 
we should emphasize that this method cannot not lead to any artificial 
prolongation of the stage of inflation. At the stage of  self-reproduction 
of the universe the number of new independent domains of the size 
$H^{-1}$ created due to quantum fluctuations and expansion of the 
universe is much greater than the number of domains disappearing 
at the boundaries.}
If the probability distribution is not stationary, we follow development of
our distribution at every  step of our calculations. However, if the
distributions become stationary, one can obtain much better statistics by
integrating the distribution beginning from the moment when it approaches
stationary regime.

2) {\bf The Planck boundary.}   Here we may impose different
boundary conditions, depending on our assumption concerning the
Planck-scale physics. Fortunately, the results which we obtain
are not terribly sensitive to these assumptions.

The simplest condition is to discard all points which jump over
the Planck boundary, and to renormalize the probability
distribution $P_p$ in the same way as we are doing when the
points jump over the boundary where inflation ends.

3) {\bf The boundary at large $\sigma$}.   In order to
obtain a stationary solution we may need to have an additional
boundary at large $\sigma$.  We will assume that there exists a
boundary at some sufficiently large value of the field $\sigma =
\sigma_b$. For simplicity, we will impose the same condition at
this boundary as at all other boundaries: we discard all points
which jump over this boundary, and renormalize the probability
distribution $P_p$ after such jumps occur.

An alternative possibility is to consider that the effective
potential $V(\sigma)$ becomes very steep at large $\sigma$, and
the distribution $P_p$ becomes stationary without any need for
imposing additional boundary conditions at large $\sigma$.

To plot the distribution we make a two-dimensional histogram of
$\sigma$ and $\phi $ and fill the histogram with the points
corresponding to each new step of our calculations. After  the
distribution approaches a stationary regime, we instead make a
histogram which includes all points starting with the step at
which the distribution became almost stationary. This does not
change the shape of the stationary distribution, but effectively
increases the number of points involved in the calculation, and
decreases relative deviation of our `experimental results' from
the   probability distribution $P_p$. After several hundred more
steps we get a rather smooth picture of the stationary
distribution $P_p(\sigma,\phi)$.

In this paper we will present the results of our simulations and
of analytic investigation for potentials $V(\sigma)$ of several
different types.  The results of our calculations will be
represented as a distribution $P_p$ inside a box with axes
$x$ and $y$ corresponding to the values of the fields $\sigma$
and $\phi$. The field $\sigma$ grows along the $x$-axis from
$\sigma = 0$ in the left lower corner. The field $\phi$ grows
from $\phi = 0$ when one goes upwards along the $y$-axis from
the same corner.  The height  $z$ of the surface in the box
will correspond  to the value of $P_p(\sigma,\phi;t)$.  We will
not make any attempt to make our calculations with realistically
small or large values of parameters; our purpose is just to
present   the most important qualitative features of the
distribution $P_p$.

\section{\label{Example} Stochastic Processes in Brans--Dicke
Theory with a Constant Vacuum Energy Density}
In order to get some insight into the complicated behavior of
two fluctuating scalar fields, $\sigma$ and $\phi$, we will
temporarily make two steps back to simplify our model. First of
all, we will consider the theory with the simplest effective
potential  $V(\sigma) = V_0 = const$.  Also, we will return for
a moment from the Brans--Dicke theory to the standard Einstein
theory. This is equivalent to keeping the field $\phi = M_{\rm
p}\sqrt{\omega\over\ {2\pi}}$ fixed. In this case $H^2 =
H_0^2 \equiv {8\pi V_0\over3 M_{\rm p}^2}= const$. The diffusion
equation for $P_p$ in this theory looks very simple,
\begin{equation}\label{phyy}
{\partial P_p(\sigma,t)\over\partial t} = {H_0^3\over 8\pi^2}\,
{\partial^2P_p(\sigma,t)\over\partial\sigma^2}+3H_0 P_p(\sigma,t)\ .
\end{equation}
The solution to the diffusion equation in the comoving frame is
a Gaussian with increasing dispersion $\Delta^2(t) = {H_0^3\over
4\pi^2}\ t$. Since the potential is constant, diffusion will
always dominate classical motion and the Universe will be
eternally self-regenerating, with a probability distribution in
the physical frame $\ P_p(\sigma,t) \sim \exp(3H_0t)\
\exp\Bigl(-{(\sigma - \sigma_o)^2\over 2\Delta^2(t)}\Bigr)$
that grows exponentially with time, due to the increase in the
physical volume of the Universe.

Let us now consider the same constant potential $V(\sigma) =
V_0$ in the Brans--Dicke theory. The classical equations of
motion read
\begin{equation}\label{CST}
\dot{\sigma} = 0\ , \hspace{5mm} \dot{\phi} =
\left({4V_0\over3\omega}\right)^{1/2}
= const\ .
\end{equation}
As before, quantum diffusion of the field $\sigma$ dominates its
classical motion, and the Universe will be in the stage of
eternal self--reproduction for the field  $\phi$ in the interval
\begin{equation}\label{ESR}
\phi_p\equiv\left({V_0\omega^2\over4\pi^2}\right)^{1/4} <\, \phi \,
< \left({V_0\omega^3\over3\pi^2}\right)^{1/4} \ .
\end{equation}

Let us now analyze the behavior of the probability distribution
$P_p$ when the field $\phi$ enters the self--reproduction range
(\ref{ESR}). Since $H \propto 1/\phi$, the distribution will
move towards the Planck boundary $\phi = \phi_p$ (\ref{ESR})
along the $\phi$--direction.  Very soon the distribution
$P_p$ reaches the Planck boundary and remains concentrated in a
very narrow region near it, see Fig. 1. As a result, the only
possible changes of the distribution $P_p$ become related to the
Brownian motion of the field $\sigma$ along the Planck boundary.
The jumps of this field along the Planck boundary are
proportional to $H_p/2\pi$, where $H_p$ is the Hubble constant
near the Planck boundary, $H^2_p = {8\pi\over3}\ V_0^{1/2}$.
This suggests (and the results of our computer simulations
apparently confirm this conjecture) that the behavior of $P_p$
at large $t$ can be approximately described by equation for
one-dimensional diffusion along the Planck boundary, similar to
eq. (\ref{phyy}):
\begin{equation}\label{PHY}
{\partial P_p\over\partial t} = {H_p^3\over8\pi^2}\,
{\partial^2P_p\over\partial\sigma^2} + 3H_p P_p\ .
\end{equation}
The solution to (\ref{PHY}) is, like in general relativity, a
Gaussian with increasing dispersion $\Delta^2(t) = {H_p^3\over
4\pi^2}\ t$ and an exponentially growing factor that accounts
for the increase in volume at the Planck boundary,
\begin{equation}\label{RUN}
\ P_p(\sigma,t) \sim  \exp(3H_pt)\ \exp\left(-{(\sigma
- \sigma_o)^2\over 2\Delta^2(t)}\right) \ .
\end{equation}
It is clear that in this case the distribution gradually
becomes 
flat everywhere along the Planck boundary. Therefore the main
part of the volume of the Universe will be in a state with
indefinitely large $\sigma$. However, as soon as the dispersion
$\Delta(t)$ becomes greater than the distance between $\sigma_0$
and $\sigma$, the ratio  of the volume occupied with a field
$\sigma$ to the volume containing field $\sigma_0$ becomes
time-independent.  This is an example of the runaway
stationarity that we described in the Introduction. It is
essential that the potential be sufficiently flat in order to
have this kind of stationarity.  For example, such a regime may
occur in the theories with $V(\sigma) \sim V_0 (1 -
e^{-\sigma^2})$.

On the other hand, a potential $V(\sigma) \sim \sigma^{2n}$
will not present  this runaway stationarity, since any
$\sigma$--dependence in $H \propto\sigma^n/\phi$ will make the
distribution move forever towards large values of $\sigma$ until
it is strongly peaked at infinity, unless we impose an extra
boundary condition at $\sigma=\sigma_b$. However, the behavior
of the distribution along the $\phi$--direction will follow the
same pattern as above, being rapidly concentrated along the
Planck boundary (in general a complicated curve, depending on
the form of the potential $V(\sigma)$). Its subsequent evolution
will be reduced to essentially a one--dimensional diffusion
along the Planck boundary.

\begin{figure}[h!]
\label{KKLTpot} \centering 
\includegraphics[scale=1.7]{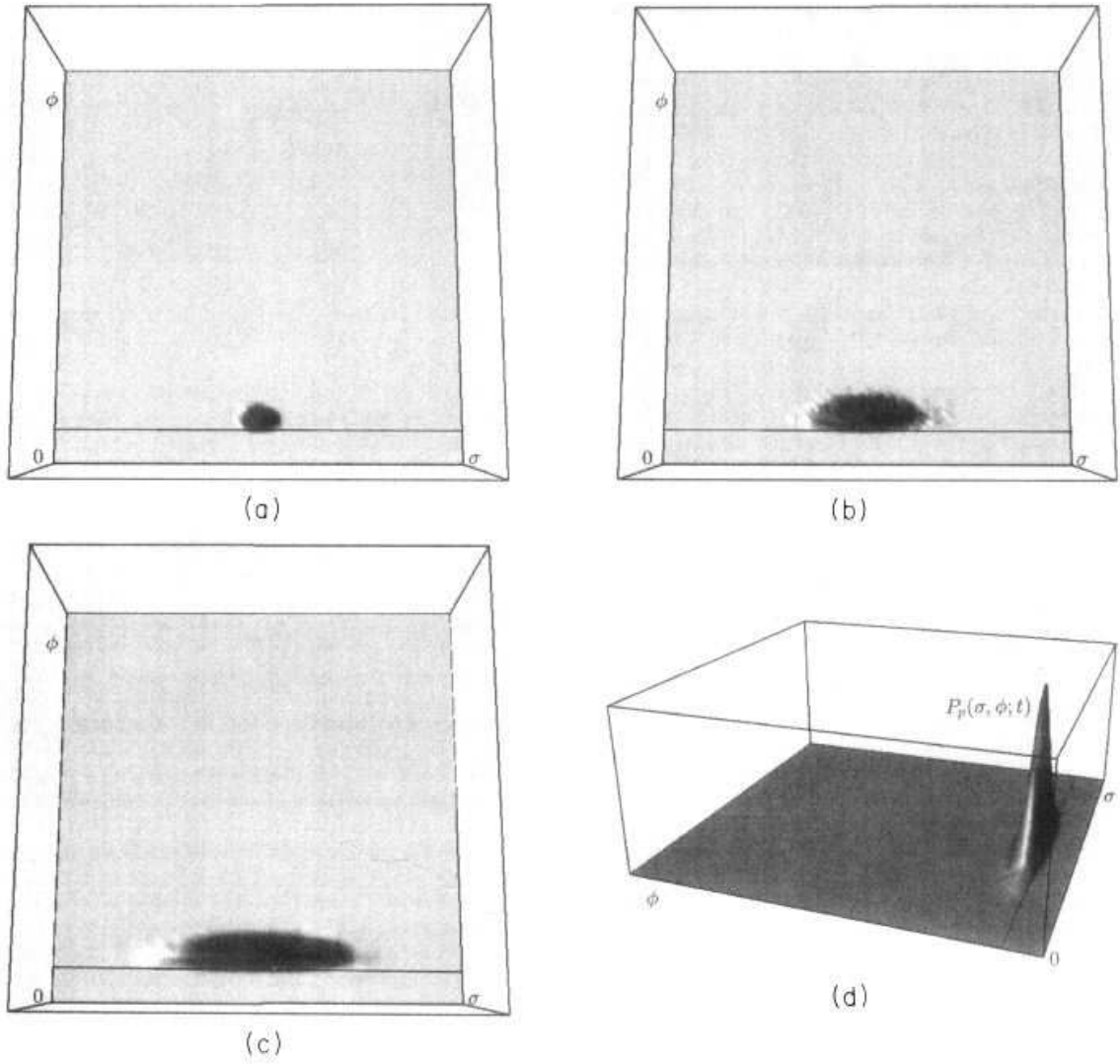} \caption{\small Diffusion of the distribution $P_p$ along the
Planck boundary in the simplest theory with $V(\sigma) = const$,
in the time $t$ parametrization. Figs. 1a, 1b, 1c show different
steps towards stationarity at the Planck boundary.
Fig. 1d shows the same stage of the diffusion as
Fig. 1c, but in a different perspective.}\label{fig1} 
\end{figure}

To  illustrate  this feature, we performed computer simulation
of diffusion for the simplest case of the theory with $V(\sigma)
= const$, see Fig. 1.  The first few images (Figs. 1a - 1c)
correspond to the view  ``from the top''. The horizontal line
across the box  in Figs. 1a - 1c corresponds to the Planck
boundary; the distribution $P_p$ is concentrated above this
line. In the beginning we have a delta-functional  distribution
concentrated near some initial values of $\sigma$ and $\phi$.
Then it rapidly moves towards the Planck boundary; it looks like
a round spot from the viewpoint we have chosen, see Fig. 1a.
After that the distribution widens in the $\sigma$-direction
along the Planck boundary, while preserving its width in the
$\phi$-direction orthogonal to this boundary, see Figs 1b, 1c.
Its shape can be better understood from another viewpoint, see
Fig. 1d, which  shows the same distribution as Fig. 1c in a
different perspective. The most important feature of this
distribution is that its evolution very soon becomes effectively
one-dimensional, being entirely concentrated near the Planck
boundary.  We will take advantage of this feature for the study
of the runaway solutions in the next Section.

\section{\label{run}Runaway Solutions}
Let us study the behavior of the probability distribution $P_p$
along the Planck boundary. In the case of an increasing
potential like $\lambda \sigma^{2n}$, the larger the value of
$\sigma$ in a given domain, the greater the increase in physical
volume of that domain.  Therefore, the probability distribution
$P_p$ will tend to move  towards large $\sigma$. The way it
moves will depend on the type of potential. For some, as we will
see, it is an explosive behavior. The probability distribution
gives a statistical description of the quantum diffusion process
towards large $\sigma$, but it proves useful to analyze the
particular behavior of those relatively rare domains in which
the field $\sigma$ increases in every quantum jump of amplitude
$H/2\pi$.  We can compute the speed at which those domains move
towards large values of the field $\sigma$ from the equation
(where $\Delta t = H^{-1}$)
\begin{equation}\label{STH}
\dot{\sigma} \equiv {\delta\sigma\over\Delta t} = {H^2_p\over2\pi}
= {4\over3}\left({\lambda\over2n}\right)^{1/2} \sigma^n\ ,
\end{equation}
where $H_p$ is the Hubble parameter along the Planck boundary.
For $n=1$ there is an exponential increase of $\sigma$ in those
domains, while for $n>1$ we find an explosive solution
\begin{equation}\label{SST}
\sigma(t)=\sigma_o\ \left(1-{4(n-1)\over3} \sqrt{\lambda\over2n}
\ \sigma_o^{n-1}\ t\right)^{-{1\over n-1}}\ .
\end{equation}
We see that for all $n>1$ those first domains of the diffusion
process reach infinity in finite time. Note that the total
volume of such domains at that time will be finite, and then
they will start growing at an infinitely large rate.  This
behavior is explosive and will correspond to probability
distributions that are nonstationary and singular at $\sigma \to
\infty$.  It is extremely difficult to study this regime
using computer simulations, but with the help of the results
obtained in the previous section we can get a pretty good
understanding of the behavior  of $P_p$ in such a situation.

The qualitative analysis and the computer simulations of Section
\ref{4} suggest that the general solution to the diffusion
equation for both fields will factorize naturally into a motion
perpendicular to the Planck boundary  that will reach
stationarity very quickly, and a motion along it towards large
values of $\sigma$.  We will try to study this last
motion in the absence of a boundary condition at $\sigma_b$. We
assume that we can ignore the classical motion along the Planck
boundary. Indeed, for large $\omega$ and small masses and
coupling constants classical motion is almost exactly orthogonal
to the Planck boundary. This assumption might be violated at
large $\sigma$ for some theories with very rapidly growing
potentials, but in the most interesting case of the theory
$\lambda\sigma^4$ the classical motion is exactly orthogonal to
the Planck boundary for all $\sigma$. On the other hand, for
large $\omega$ and small masses and coupling constants the
motion along the Planck boundary almost exactly coincides with
the motion along the $\sigma$-axis. In this case an approximate
diffusion equation for $P_p$ analogous to eq. (\ref{PHY}) can
then be written as follows:
\begin{equation}\label{RFK}
{\partial P_p\over\partial t} = {1\over8\pi^2}
{\partial\over\partial\sigma}\left(H_p^{3/2}
{\partial\over\partial\sigma}\left(H_p^{3/2} P_p\right)\right)
+ 3H_p P_p\ ,
\end{equation}
where the Hubble parameter along the Planck boundary in the
theories with $V = {\lambda\over 2n}\sigma^{2n}$ is given by
\begin{equation}\label{RHS}
H_p^2 = {8\pi\over3} {V(\sigma)\over M_{\rm p}^2(\phi)} =
{8\pi\over3} V(\sigma)^{1/2} = {8\pi\over3}\left({\lambda\over2n}
\right)^{1/2} \sigma^n \ .
\end{equation}
Equation (\ref{RFK}) can be written as
\begin{equation}\label{PUS}
{\partial\Psi\over\partial u} =
{\partial^2\Psi\over\partial s^2} + a\ s^p \Psi\ ,
\end{equation}
where
\begin{equation}\begin{array}{c}\label{PUC}
{\displaystyle  
\Psi = \exp(Eu) H^{3/2} P_p\ , \hspace{5mm}\hspace{5mm}
u=\sqrt{8\over27\pi} \left({\lambda\over2n}\right)^{3/4}
\left({4-3n\over4}\right)^2 t\ , }\\[3mm]
{\displaystyle
s = \sigma^{4-3n\over4}\ , \hspace{5mm}\hspace{5mm} a = 9\pi
\left({4\over4-3n}\right)^2
\sqrt{2n\over\lambda}\ , \hspace{5mm}\hspace{5mm} p =
{2n\over4-3n}\ .}
\end{array}\end{equation}

Let us first analyze the case $V(\sigma) = \lambda \sigma^4/4$.
This corresponds to $p=-2$. It is clear from (\ref{PUS}) that
there are no stationary solutions to this equation in the
absence of a boundary condition for $\sigma$.  Furthermore, we
know from quantum mechanics that potentials of the type $-1/s^2$
  have singular solutions at $s=0$ ($\sigma = \infty$).
Therefore we expect the distribution $P_p$ to be singular at
$\sigma=\infty$.  This was also expected from the analysis of
those first domains that reach infinity in finite time.

On the other hand, for the theory $V(\sigma)=m^2 \sigma^2/2$,
the Schr\"odinger potential (\ref{PUS}) is of the type $-s^2$,
which has non singular solutions at large $\sigma$. For small
$s$ the solution of eq. (\ref{PUS}) is a Gaussian centered at
$s(t) = \left\langle s\right\rangle$, with dispersion
$\Delta(t)$. The potential then acts asymmetrically on it,
pulling more at large $s$ and leaving a long exponential tail at
small $s$. The maximum will move towards large $\sigma$, while
maintaining a regular solution at infinity. This is expected
from the previous analysis of the most rapid domains
(\ref{STH}). For $n=1$ we find that the first domains will take
an infinite time to reach infinity, giving a regular solution at
any time.

However, in both cases we do not have runaway stationarity.
Indeed, let us consider equation (\ref{RFK}) and assume (for
simplicity only) that in the very beginning the function $P_p$
was constant. Then both for $n = 1$ and for $n = 2$ the first
term in the r.h.s. of this equation initially is positive.
Neglecting this term, we obtain $P_p \sim \exp(3H(\sigma)t)$,
which does not exhibit any runaway stationarity. Taking into
account the first term in the r.h.s. of  equation (\ref{RFK})
makes the growth of $P_p$ at large $\sigma$ even faster.  This
confirms our expectations that in order to obtain runaway
stationarity one should have an extremely flat effective
potential.

\section{\label{stat}Stationary Distributions for Various Theories}
In this section we will study  stationary probability
distributions for several different potentials $V(\sigma)$. Our
investigation will mainly rely on the results of our computer
simulations, but we will try to make analytical investigation
whenever possible.  As we have argued in the previous sections,
the simultaneous diffusion of both fields can be approximated by
a quick diffusion in the $\phi$--direction towards the Planck
boundary and a subsequent diffusion along it until it reaches
the boundary at $\sigma = \sigma_b$, or until a stationary
distribution is established for some other reason.

1) $V(\sigma) = {\lambda\over 4}\,\sigma^4$.  In this case the
motion along the Planck boundary is governed by (\ref{PUS}) with
$p=-2$ and $a=72\pi/\sqrt\lambda$ and with the boundary
conditions $\Psi(0) = \Psi(s_b) = 0$. It is difficult to find an
exact analytical solution to this simple equation, but one can
easily solve it in the WKB approximation,
\begin{equation}\begin{array}{c}\label{WKB4}
{\displaystyle
\Psi(s)\sim\left(E - {a\over s^2}\right)^{-1/4}\exp\left[
-(Es^2-a)^{1/2} + \sqrt{a}\ {\rm arcsec}\sqrt{Es^2\over a}\,\right],
\hspace{5mm} Es^2 > a,  }\\[4mm]
{\displaystyle
\Psi(s)\sim 2\left({a\over s^2} - E\right)^{-1/4}\cos\left[
(a-Es^2)^{1/2} - \sqrt{a}\ \ln\left({\sqrt a + (a-Es^2)^{1/2}\over
s\sqrt E}\right) + {\pi\over4} \right], \hspace{5mm} Es^2 < a,  }
\end{array}\end{equation}
where $E \simeq {72\pi\sigma_b\over\sqrt\lambda}\
(1-{3\over4}(\pi\sqrt\lambda/6)^{1/3})$.  This solution has a
very sharp maximum close to the boundary $\sigma= \sigma_b$
and an exponential decay for small $\sigma$ (large $s^2=1/\sigma$).
This behavior is precisely what we observe in the numerical
solutions described below.

We take the following parameters for our computer simulations:
$\omega = 50$, $\lambda = 0.3$. In the beginning of the series
of calculations we took the points with coordinates
$(\sigma,\phi)$ close to the Planck boundary, but in different
initial positions with respect to the boundary $\sigma_b$. We
have found that the duration of the intermediate non-stationary
regime depends on the initial values of $\phi$ and $\sigma$.
However, typically the stationary distribution
$P_p(\sigma,\phi)$ is established very rapidly, after just a few
steps $\Delta t \sim H^{-1}_0$.  The resulting stationary
distribution is presented in Fig. 2. Line OA at this figure
corresponds to  the end of inflation; there is no inflation for
($\sigma, \phi$) to the left of this line.  Line OB corresponds
to the Planck boundary; $V(\sigma)$ is greater than the Planck
density $M_{\rm p}(\phi)$ under this line. The line $\sigma =
\sigma_b$ is an additional boundary discussed above.

\begin{figure}[h!]
\label{KKLTpot} \centering 
\includegraphics[scale=1.8]{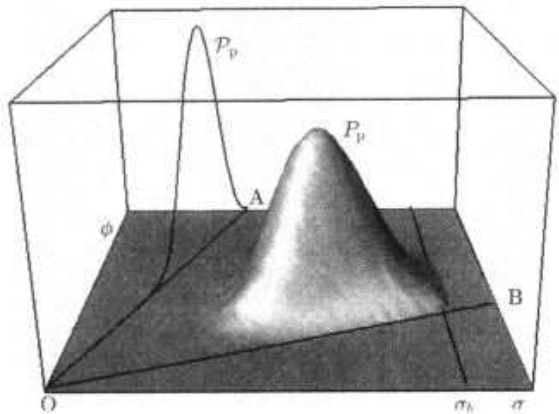} \caption{\small Stationary probability distribution $P_p$ in the plane
$(\sigma, \phi)$ for the theory $V(\sigma) = {\lambda\over4}
\sigma^4$, in the time $t$ parametrization.  The line OA
corresponds to the end of inflation, while the line OB to the
Planck boundary.  Inflation occurs between these two lines.  The
line above OA shows the probability distribution ${\cal P}_p$.}\label{fig2} 
\end{figure}

The curve above the line OA is of the most interest for us. It
represents the stationary flow of domains crossing different
parts of the boundary OA at the end of inflation. This
gives us the probability distribution ${\cal P}_p(\phi_e)$ that
at the end of inflation the field $\phi$ takes some particular
final value $\phi_e$. The maximum of this curve corresponds to
the most probable value of the Planck mass $M_{\rm p} =
\sqrt{2\pi\over \omega} \phi_e$ at the end of inflation. The top
of the `mountain', which shows the stationary distribution
$P_p(\sigma,\phi)$,  corresponds to the most probable value of
$M_{\rm p}$ during inflation. Not unexpectedly, the curve above
the line OA looks like a shadow of the mountain $P_p$. Indeed,
as we have already mentioned, the distribution $P_p$ and the
distribution ${\cal P}_p$ are directly related to each other,
see (\ref{PFP}). Meanwhile, the shape of the distribution $P_p$
at small $\sigma$ and large $\phi$ is obviously related to its
shape at large $\sigma$ and small $\phi$, since in the
intermediate regions the points ($\phi, \sigma$) follow
classical circular trajectories, see e.g.  (\ref{CLA}).
Consequently, the position of the maximum of ${\cal P}_p$ can be
approximately obtained by drawing a circle with the center at
$\phi= \sigma = 0$, which goes through the crossing point of the
Planck boundary and the boundary $\sigma =
\sigma_b$.

2) $V(\sigma) = {\lambda \sigma^4\over 4}\,
\exp{\sigma^4\over\sigma_0^4}$. The exponential term is added
here in order to show that one can avoid introducing additional
boundaries at $\sigma = \sigma_b$ if the effective potential
becomes very steep at large $\sigma$.  The result is that the
WKB solution (\ref{WKB4}), instead of decreasing sharply to
$\sigma=\sigma_b$, decays exponentially fast. It is just the
effect of substituting an infinite barrier by an exponential
barrier.  As we see in the numerical simulations, with a proper
choice of the place where the effective potential becomes very
steep one can reproduce the same result as if there were a
boundary at $\sigma_b \sim \sigma_0$, see Fig. 3.

\begin{figure}[h!]
\label{KKLTpot} \centering 
\includegraphics[scale=1.8]{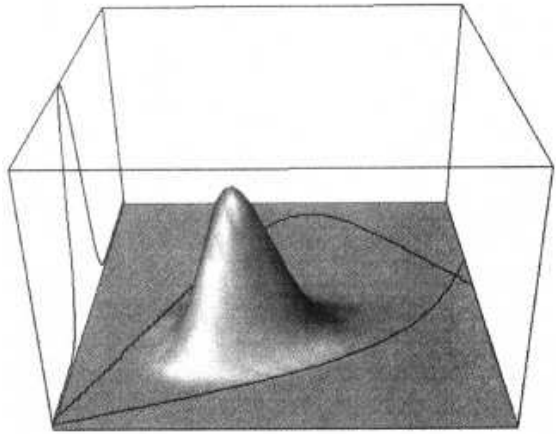} \caption{\small Same as in Fig. 2 for the potential $V(\sigma) =
{\lambda \sigma^4\over 4}\,  \exp{\sigma^4\over\sigma_0^4}$.
The distribution ${\cal P}_p$ is shown on the left side of the box.}\label{fig3} 
\end{figure}

In this figure we show the boundary of the end of inflation by a
somewhat wavy line above the Planck boundary. At small $\sigma$
the boundary of the inflationary region goes as a straight line
from the point $\sigma = 0$, $\phi = 0$ (compare to the line OA,
Fig. 2), but then it becomes curved because of the exponential
term which precludes inflation at large $\sigma$.  Finally this
line crosses the Planck boundary. The distribution $P_p$ is
surrounded by this line and the Planck boundary.

On the left wall of the box we show the distribution ${\cal
P}_p$, which in the previous picture we have shown above the
line OA. In other figures we will do the same everywhere when
the end of inflation boundary is significantly curved.

3) $V(\sigma) = {\lambda \sigma^4\over 4}\,
\exp(-{\sigma^4\over\sigma_0^4})$. In this case there is a sharp
cut-off of the effective potential at $\sigma > \sigma_0$, which
also leads to the existence of a stationary solution, as if
there were a boundary near $\sigma_0$, see Fig.  4. As in Fig.
3, the wavy line corresponds to the end of inflation boundary.

\begin{figure}[h!]
\label{KKLTpot} \centering 
\includegraphics[scale=1.8]{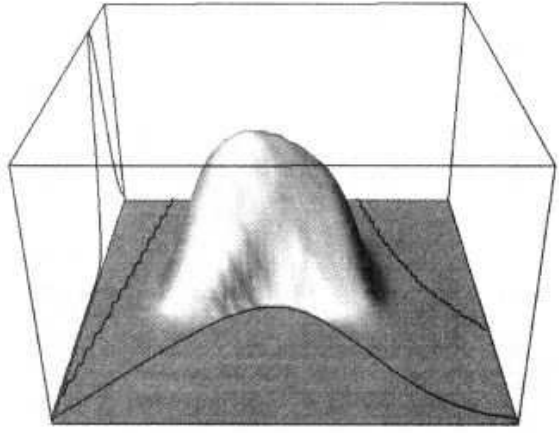} \caption{\small Same as in Fig. 1 for the potential $V(\sigma) =
{m^2\over 2}\sigma^2$.}\label{fig4} 
\end{figure}

4) $V(\sigma) = {m^2\over 2}\sigma^2$. In this case the Planck
boundary is not a straight line but a parabola $\phi^2 = {m\
\omega\over2\pi\sqrt2}\ \sigma$. The eigenvalue equation
associated with the probability distribution along the Planck
boundary is (\ref{PUS}) with $p=2$ and $a=144\pi\sqrt2/m$.
In the WKB approximation
\begin{equation}\begin{array}{c}\label{WKB2}
{\displaystyle
\Psi(s)\sim\left(E - as^2\right)^{-1/4}\exp\left[{s\over2}
(E-as^2)^{1/2} + {E\over2\sqrt a}\arcsin\sqrt{as^2\over E}
- {\pi E\over4\sqrt a}\right], \hspace{5mm} as^2 < E\ ,  }\\[3mm]
{\displaystyle
\Psi(s)\sim 2\left(as^2 - E\right)^{-1/4}\cos\left[{s\over2}
(as^2-E)^{1/2} - {E\over2\sqrt a}\ln\left({s\sqrt a +
(as^2-E)^{1/2}\over
\sqrt E}\right) - {\pi\over4} \right] , \hspace{3mm} as^2 > E\ ,}
\end{array}\end{equation}
where $E\simeq {144\pi\sqrt{2\sigma_b}\over m}\ (1- ({9\pi m
\over16\sqrt2\sigma_b})^{1/3})$.  This solution has a very
sharp maximum close to the boundary $\sigma= \sigma_b$ and an
exponential decay for small $\sigma$ (small $s^4=\sigma$).  This
behavior is precisely what we observe in the numerical
solutions.  The distribution $P_p$ is very similar to that of
the theory ${\lambda\over 4}\ \sigma^4$, see Fig. 5.

\begin{figure}[h!]
\label{KKLTpot} \centering 
\includegraphics[scale=1.8]{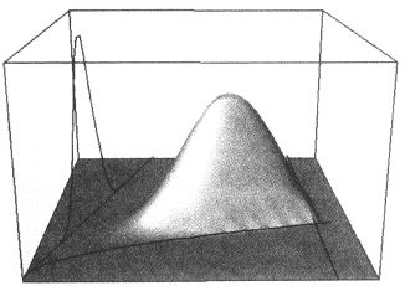} \caption{\small Same as in Fig. 3 for the potential $V(\sigma) =
{\lambda \sigma^4\over 4} \,
\exp\Bigl(-{\sigma^4\over\sigma_0^4}\Bigr)$.}\label{fig5} 
\end{figure}

5)  $V(\sigma) = {m^2\over 2}\,\sigma^2 + {\lambda\over
4}\sigma^4 \log{\sigma\over\sigma_0}$. Naively, one could expect
that the main part of the volume of the Universe in this model
should originate as a result of inflation beginning from the
points on the Planck boundary with the smallest angle $\theta$.
Indeed, eq. (\ref{CLA}) shows that the smaller is the initial
angle, the greater is the degree of inflation \cite{ExtChaot}.
However, due to the self-reproduction of inflationary domains
and the more rapid expansion of domains with greater $\phi$
along the Planck boundary, the distribution $P_p$ does not stay
near the point with the smallest $\theta$, but moves towards the
largest possible $\phi$ and $\sigma$, see Fig. 6. As we
discussed in the last section, this is a general result for any
increasing potential, so $P_p$ is expected to have a maximum close
to the boundary $\sigma=\sigma_b$.

\begin{figure}[h!]
\label{KKLTpot} \centering 
\includegraphics[scale=1.8]{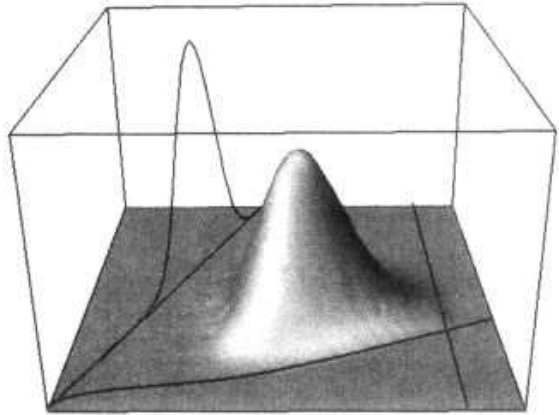} \caption{\small Same as in Fig. 1 for the potential $V(\sigma) =
{m^2\over 2}\,\sigma^2 + {\lambda\over 4}\sigma^4
\log{\sigma\over\sigma_0}$.}\label{fig6} 
\end{figure}

6)  $V(\sigma) = {1\over 4\lambda}(m^2- {\lambda\sigma^2})^2$.
This is a typical potential used in the theories with
spontaneous symmetry breaking. It has a minimum at $\sigma_0 =
{m\over \sqrt\lambda}$. In this theory we have two alternative
regimes.  If one begins at $\sigma = 0$, one may have an
inflationary regime at small $\sigma$ \cite{Hybrid}, similar to
the inflationary regime in the new inflationary universe
scenario. If, on the other hand, inflation begins at large
$\sigma$, then one has an inflationary regime similar to that in
the theory ${m^2\over 2}\,\sigma^2 + {\lambda\over 4}\sigma^4$.
The only difference is that in the theory under consideration
the field $\sigma$ eventually rolls down not to $\sigma = 0$,
but to $\sigma =\sigma_0$. The first possibility is illustrated
by Fig. 7.  On this figure the point $\sigma = 0$ corresponds
not to the left corner, as usual, but to the center of the
$x$-axis. 

\begin{figure}[h!]
\label{KKLTpot} \centering 
\includegraphics[scale=1.8]{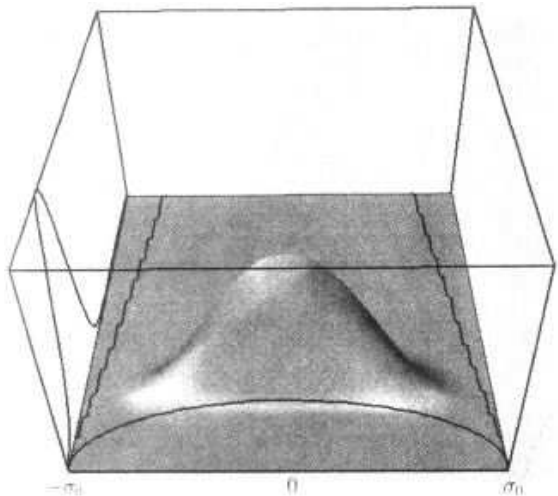} \caption{\small Same as in Fig. 1 for the potential $V(\sigma) =
{1\over 4\lambda}(m^2- {\lambda\sigma^2})^2$. It describes the
stationary distribution for the case that inflation begins at
$\sigma < \sigma_o$.}\label{fig6} 
\end{figure}

Note that if one begins with several inflationary
domains with different initial conditions, at small $\sigma$ and
at large $\sigma$, the domains with large $\sigma$ always win,
and the distribution $P_p$ very soon becomes almost entirely
concentrated at $\sigma > \sigma_0$, see Fig. 8.

\begin{figure}[h!]
\label{KKLTpot} \centering 
\includegraphics[scale=1.7]{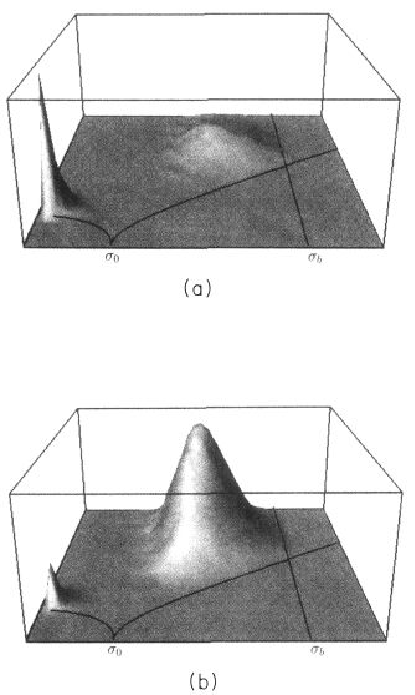} \caption{\small  Same as in Fig. 7  
for initial conditions both at
large and small $\sigma$. This series of pictures shows that if
one begins with equal number of  ``points''  at large and at
small $\sigma$, the volume corresponding to large $\sigma$
 always dominates in the limit $t \to \infty$.}\label{fig6} 
\end{figure}

\section{\label{6} $\tau$--parametrization}
The usual Fokker-Planck equation is written in terms of a time
parameter $t$ as measured by the synchronized clocks of comoving
observers.  However, in general relativity one can use many
different time parametrizations. For example, one can measure
time by the local growth of the scale factor of the Universe and
define a new time parameter \cite{b60}
\begin{equation}\label{TAU}
\tau = \ln \frac{a(x,t)}{a(x,0)} = \int_0^t dt'\
H(\sigma(x,t'),\phi(x,t')) \ .
\end{equation}
This `time' proves to be rather convenient since in this time,
by definition, all parts of the Universe expand with the same
speed $\sim e^\tau$, and $P_p$ is proportional (though not equal
\cite{LLM})
to $P_c$.\footnote{Note that $dt = H^{-1}(\sigma,\phi)\,d\tau$.
Therefore if one is interested in the probability distribution
over the invariant four-dimensional volume, one should take into
account the corresponding sub-exponential   corrections to
the leading $e^{3\tau}$ dependence of the 3D-volume on the time
$\tau$.} One can easily derive the classical equations of motion
and quantum diffusion in the new parametrization,
\begin{equation}\label{TEQ}
\frac{\partial\phi}{\partial \tau} = - \frac{M_{\rm
p}^2(\phi)}{2\pi H}\ \frac{\partial H}{\partial \phi} \ , \ \ \  \
\frac{\partial\sigma}{\partial \tau} = - \frac{M_{\rm p}^2
(\phi)}{4\pi H} \ \frac{\partial H}{\partial\sigma} \ ,
\end{equation}
\begin{equation}\label{CLT}
\frac{\partial}{\partial \tau}\left\langle\phi^2\right\rangle=
\frac{  H^2} {4\pi^2} +
{2\over\omega}\left\langle\phi^2\right\rangle \ , \ \ \ \
\frac{\partial}{\partial \tau}\left\langle\sigma^2\right\rangle=
\frac{  H^2}{4\pi^2} - {2m^2\over3H^2} \left\langle\sigma^2
\right\rangle  \ .
\end{equation}

The conditions for the self-reproduction of the inflationary
universe are the same as in the time $t$ parametrization
(\ref{BIF}). Furthermore, we are interested in the diffusion
equation for the scalar fields in the physical frame, where
$P_p(\sigma,\phi;\tau) = P_c(\sigma,\phi;\tau) \cdot e^{3\tau}$
satisfies
\begin{equation}\begin{array}{rl}\label{TPL}
{\displaystyle
\frac{\partial P_p}{\partial \tau} =}&{\displaystyle
\frac{\partial}{\partial\sigma}\left(\frac{M_{\rm p}^2(\phi)}{4\pi H}
\frac{\partial H}{\partial\sigma} P_p + \frac{H}{4\pi}
\frac{\partial}{\partial\sigma} \left(\frac{H}{2\pi} P_p \right)
\right) } \\[3mm]
 +&{\displaystyle
\frac{\partial}{\partial\phi}\left(\frac{M_{\rm p}^2(\phi)}{2\pi H}
\frac{\partial H}{\partial\phi} P_p + \frac{H}{4\pi}
\frac{\partial}{\partial\phi} \left(\frac{H}{2\pi} P_p \right)
\right) + 3 P_p\ .}
\end{array}\end{equation}
In this case, thanks to the absence of the $3HP_p$ term, there
is a candidate for an  exact stationary solution of eq.
(\ref{TPL}) given by
\begin{equation}\label{PTT}
P_p(\sigma,\phi;\tau) \propto e^{3 \tau}\, H^{-1}(\sigma,\phi)
\exp\left({3M_{\rm p}^4(\phi)\over8 V(\sigma)}\right)\ .
\end{equation}
Note that this expression is proportional to the square of the
Hartle--Hawking wave function of the Universe.  Unfortunately,
this `stationary solution' does not actually exist in any
realistic model of inflation for the reason explained in
\cite{LLM}: The maximum of this distribution coincides with the
position of the minimum of $V(\sigma)$ where there is no
inflation and our stochastic equations do not apply. To find a
correct solution, we should impose boundary conditions
\cite{LLM},
\begin{equation}\begin{array}{c}\label{BBB}
{\displaystyle
\frac{\partial}{\partial\sigma} \left(H(\sigma,\phi)
P_p(\sigma,\phi)\right)|_{\rm end} = 0\ ,} \\[3mm]
{\displaystyle
P_p(\sigma,\phi)|_{\rm Planck} = 0\ ,} \\[3mm]
{\displaystyle
P_p(\sigma,\phi)|_{\sigma_{\phantom {}_b}} = 0 \ ,}
\end{array}\end{equation}
which considerably modifies the shape of the distribution $P_p$.
Nevertheless, the naive `solution' (\ref{PTT}) tells us
something important about the shape of the distribution $P_p$.
We  expect this distribution to be peaked at the end of
inflation boundary, instead of being concentrated near the
Planck boundary as in the time $t$--parametrization. The
exponent $\exp\left({3M_{\rm p}^4(\phi)\over8 V(\sigma)}\right)$
remains constant all the way along the end of inflation boundary
for the theory $\lambda\sigma^4$. Therefore in this theory one
expects the distribution to move towards  small $\phi$ and
$\sigma$, where the prefactor $H^{-1}$  in (\ref{PTT}) is
maximal. In other theories this argument does not apply, and in
general one may obtain a stationary distribution with parameters
depending not only on the boundary of the end of inflation but
also on the boundary at $\sigma_b$. All these features are
observed in the numerical simulations.

\section{\label{comtau} Computer Simulations of diffusion in time
$\tau$ }
Computer simulations in the time $\tau$ are similar to the ones
in the time $t$, but there are several important differences.

First of all, there is no need to make any splits of the
domains. The reason is that all domains in the time $\tau$
expand with the same speed, by definition of this time as a
logarithm of expansion.  Each step of our calculation now
corresponds to a time change $\Delta \tau = u$, where $u$ is
some small number, $u < 1$.

As before, evolution of the fields in each domain consists of
several independent parts.  Each field evolves according to
classical equations of motion, (\ref{TEQ}), and in addition to
it each field makes quantum jumps by $\delta\sigma = {H \sqrt{u}
\over \sqrt 2\pi} \, \sin r_1 $,\, $ \delta\phi = {H \sqrt{u}
\over \sqrt 2\pi}\, \sin r_2$. Here $r_1$ and $r_2$ are random
numbers which are different for each point.

As in Section \ref{4}, in addition to the distribution
$P_p(\phi,\sigma,\tau)$ we computed also the volume of all
domains where inflation ended within each new time interval
$\Delta \tau$. This gives the fraction of the volume of the
Universe where inflation ends at a given time $\tau$ within a
given interval of values of the field $\phi$. We call this
distribution ${\cal P}_p(\phi_e,\tau)$. In the theories with
$V(\sigma)=\frac{\lambda}{2n} \sigma^{2n}$
\begin{equation}\label{PFP1}
{\cal P}_p(\phi_e,\tau) \sim \phi_e \cdot
P_p(\phi_e,\sigma_e,\tau) \ .
\end{equation}

It is very instructive to compare the results of computer
simulations in time $t$ and in time $\tau$ for various
effective potentials $V(\sigma)$. The general tendency we
observe is that the distribution $P_p(\phi,\sigma,\tau)$ is more
closely concentrated not near the Planck boundary or near the
boundary at large $\sigma$ (as was the case for  the
distribution $P_p(\phi,\sigma,t)$), but near the boundary
corresponding to the end of inflation. The reason for this
difference is very simple. In the time $\tau$ there is no
additional enhancement of the volume filled by the fields
corresponding to large values of the Hubble constant. Now let us
consider several particular examples.

1) $V(\sigma) = const$. This corresponds to the simple model we
considered in Section 6. In the time $t$ the distribution
rapidly moved towards the Planck boundary and then diffused
along it, Fig. 1. Evolution of $P_p$ in the time $\tau$ is quite
different, for the reason discussed above. The distribution
moves away from the Planck boundary, see Fig. 9. This explains
many features of the distribution $P_p$ in more complicated
theories to be considered below.

\begin{figure}[h!]
\label{KKLTpot} \centering 
\includegraphics[scale=1.8]{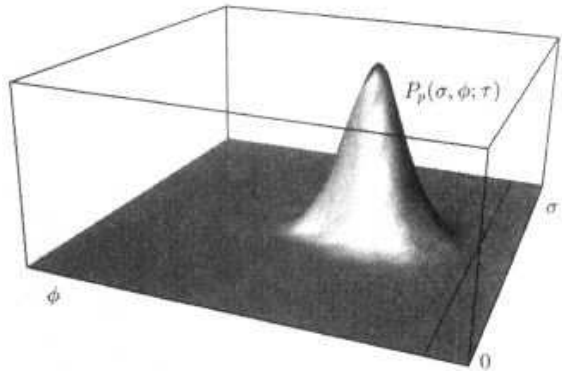} \caption{\small  In the theory with $V(\sigma) = const$ the
distribution $P_p(\phi,\sigma;\tau)$ moves away from the Planck
boundary. Compare to Fig. 1d.}\label{fig6} 
\end{figure}

2) $V(\sigma) = {\lambda\over 4}\,\sigma^4$.  In this theory we
do not have any typical mass scale which would correspond to a
maximum of the probability distribution. Therefore in time $t$
the distribution was moving towards large $\sigma$ and $\phi$,
until it was stabilized either by a boundary at large $\sigma$
or by the change of the potential at large $\sigma$ (e.g.
$V(\sigma) = {\lambda\over 4} \sigma^4\cdot \exp{\sigma^4\over
\sigma_0^4}$, see next item). In the present case the
distribution moves towards smaller and smaller $\sigma$ and
$\phi$, and there is no stationary regime unless the potential
$V(\sigma)$ at small $\sigma$ becomes, for example, quadratic in
$\sigma$, see below.  Fig. 10 shows how the distribution moves
towards the corner $\sigma = \phi = 0$. This Figure should be
compared to Fig. 2, where the corresponding distribution is
shown in time $t$.

\begin{figure}[h!]
\label{KKLTpot} \centering 
\includegraphics[scale=0.9]{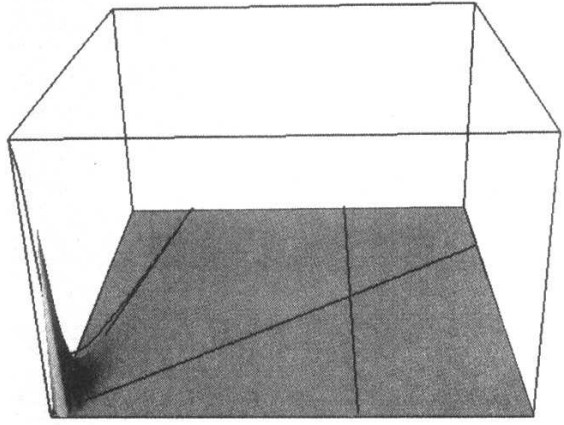} \caption{\small  Probability distribution in the plane
$(\sigma, \phi)$ for the theory $V(\sigma) = {\lambda\over4}
\sigma^4$, in the time $\tau$ parametrization. Compare to Fig. 2.}\label{fig6} 
\end{figure}

As in the Section \ref{4}, the curve above the line
corresponding to the end of inflation shows the probability
distribution ${\cal P}_p(\phi_e,\tau)$ that  at the end of inflation
the field $\phi$ takes some particular final value $\phi_e$. 
Note that in the present case this distribution depends on
 $\tau$ and moves towards $\phi_e = 0$

3) $V(\sigma) = {\lambda\over 4} \sigma^4\cdot
\exp{\sigma^4\over\sigma_0^4}$.  No qualitative difference
appears here as compared with the theory  ${\lambda\over4}\,
\sigma^4$ since the exponential term does not modify the
potential at small $\sigma$.

4) $V(\sigma) = {m^2\over 2}\sigma^2$. In this case the
distribution $P_p$ is stationary, see Fig. 11. As one can see, it
is concentrated near the boundary of the end of inflation. This
Figure should be compared to Fig. 5.

\begin{figure}[h!]
\label{KKLTpot} \centering 
\includegraphics[scale=0.9]{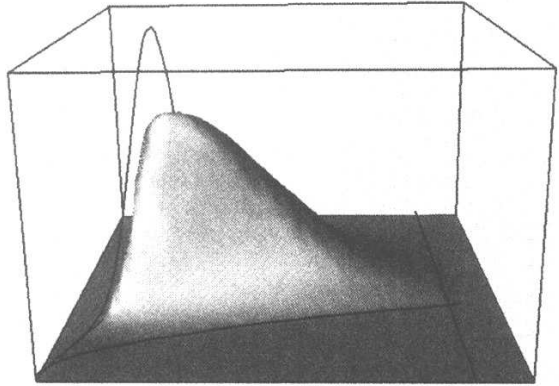} \caption{\small   Same as in Fig. 10  for the theory   $V(\sigma) =
{m^2\over 2}\sigma^2$.}\label{fig6} 
\end{figure}

5)  $V(\sigma) = {m^2\over 2}\,\sigma^2 + {\lambda\over 4}
\sigma^4 \log{\sigma\over\sigma_0}$. In this model, and in a
simpler model with $V(\sigma) = {m^2\over 2}\,\sigma^2 +
{\lambda\over 4}\sigma^4$, the quadratic term stabilizes the
distribution $P_p$, see Fig. 12.  The resulting distribution is
stationary even in the absence of the boundary at large
$\sigma$; compare with Fig. 6.

\begin{figure}[h!]
\label{KKLTpot} \centering 
\includegraphics[scale=0.9]{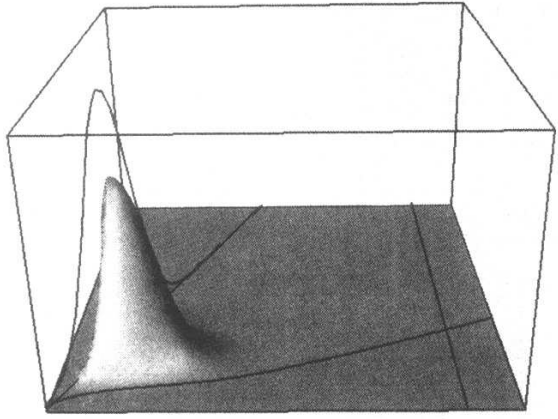} \caption{\small  Same as in Fig. 10 for the potential $V(\sigma) =
{m^2\over 2}\,\sigma^2 + {\lambda\over 4}\sigma^4
\log{\sigma\over\sigma_0}$.}\label{fig6} 
\end{figure}

6)  $V(\sigma) = {1\over 4\lambda}(m^2- {\lambda\sigma^2})^2$.
Here we have two alternative regimes.  If one begins at large
$\sigma$, then one has an inflationary regime similar to that in
the theory ${m^2\over 2}\,\sigma^2 + {\lambda\over 4}\sigma^4$.
The corresponding distribution will be very similar to that
shown in Fig. 12.  One may have an inflationary regime at small
$\sigma$ \cite{Hybrid}, similar to the inflationary regime in
the new inflationary universe scenario. The corresponding
distribution is shown in Fig. 13; compare with Fig. 7.

\begin{figure}[h!]
\label{KKLTpot} \centering 
\includegraphics[scale=0.8]{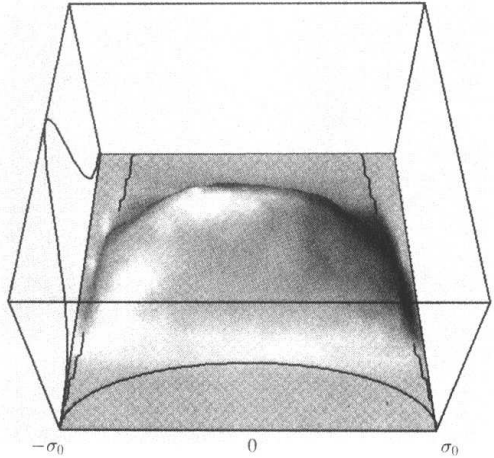} \caption{\small  Same as in Fig. 10 for the potential $V(\sigma) =
{1\over 4\lambda}(m^2- {\lambda\sigma^2})^2$. It describes the
stationary distribution for the case that inflation begins at
$\sigma < \sigma_o$.}\label{fig6} 
\end{figure}

Note that the stationary distributions we have obtained look
different from those for the same theories in time $t$. This
means that the normalized distribution $P_p$ in the large time
limit  does not depend on time, but it does depend on the choice
between different `times' ($t,\tau$, etc.) A similar conclusion
was earlier reached for other models studied in \cite{LLM}.  The
reason of this strange behavior can be understood by using the
following simple analogy.

Let us consider a two-dimensional plane $(x,y)$ and a cone
formed by the lines $y = x$ and $y = -x$ going from the point
$(0,0)$ towards positive $y$, see Fig. 14.  Let us paint light
grey the area inside this cone to the right of the $y$-axis, and
paint gray the area inside the cone to the left of the
$y$-axis.  Now we will  consider $y$ as a time direction, cross
the cone by the lines of fixed $y$ at a distance $dy$ from each
other and compare the white area and the  gray area in
the interval from $y$ to $y+dy$. Obviously, for all $dy$ the
ratio of the white area to the  gray area will be equal
to 1, independently of the time $y$ (stationarity). 

\begin{figure}[h!]
\label{KKLTpot} \centering 
\includegraphics[scale=0.4]{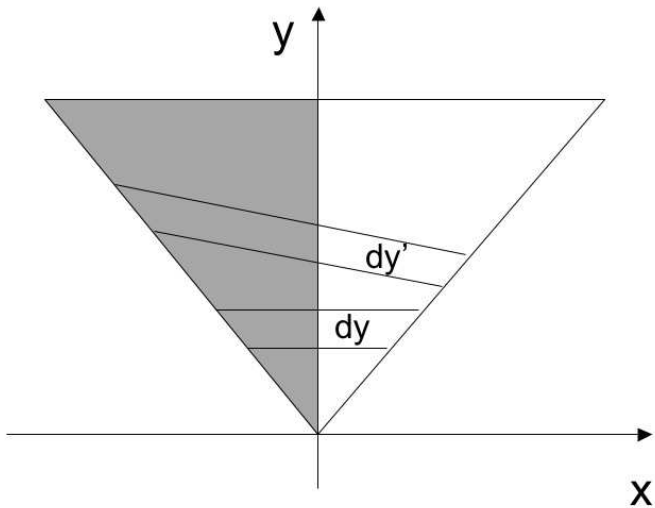} \caption{\small  A  cone in a two-dimensional space formed by the lines $y = x$ and $y = -x$. Two different
parametrizations of  ``time''  $y$ give different ratios for the
gray over the white area, although they are both
stationary.}\label{fig6} 
\end{figure}

Now one may
choose another `time' direction $y'$ by rotating the $y$-axis,
and slice the cone by the lines of constant $y'$ and $y' +dy'$.
In this case the ratio of  the white area to   the  
gray area in the interval from $y'$ to $y'+dy'$ also will not
depend on the time $y'$ (stationarity), but this ratio will not
be equal to 1 anymore. This strange effect is possible due to
the fact that the total area of the gray part of the cone,
as well as of the white one, is infinite. As usual, when
the integrals are divergent, their ratio depends on the way one
takes them.  A similar effect appears in our case as well.  The
total volume of all inflationary domains in a self-reproducing
universe is infinite in the limit $t \to \infty$ (or $\tau \to
\infty$).  Therefore the relative fraction of the volume of all
domains with any particular  properties may depend on the way in
which we are sorting out these domains.  This is the main reason
of the difference between $P_p(\sigma, \phi;t)$ and
$P_p(\sigma,\phi;\tau)$. This difference exists even if each
distribution is stationary.

\section{\label{9} Discussion}
Let us try to summarize our results and discuss their possible
implications.  First of all, we have confirmed that the regime
of self-reproduction is possible not only in the ordinary
inflationary theory, but in the Brans--Dicke inflation as well.
However, the Brans--Dicke inflation has some new interesting
features. In this theory the upper (Planck) boundary for the
energy density of a classical space-time is not a point
$V(\sigma) = M_{\rm p}^4$ as in the Einstein theory, but a line
$V(\sigma) = {4\pi^2\over \omega^2} \phi^4$. Similarly, the end
of inflation boundary in the simplest models of chaotic
inflation with $V(\sigma) \sim \sigma^n$ is not a point $\sigma
\sim M_{\rm p}$ but a line $\sigma \sim \phi/\sqrt \omega$.  In
the ordinary inflationary models the probability distribution
$P_p$ to find the inflaton field $\sigma$ at a given time in a
given volume typically is concentrated either near the Planck
boundary or near the boundary where inflation ends \cite{LLM}.
In the Brans--Dicke theory with $V(\sigma) \sim \sigma^n$ or
with $V(\sigma) \sim e^{\alpha\sigma}$ this distribution also
approaches one of these two boundaries, but after that it may
continue moving, sliding along the  boundaries.

As a result, the probability distribution $P_p$ approaches
stationary regime only if there exist  some additional reasons
which preclude this sliding. This may happen, for example,  if
the effective potential becomes very steep (or if it decreases)
at large $\sigma$. We have studied this possibility both by
making computer simulations in the theories with potentials
which become rapidly increasing (or decreasing) at large
$\sigma$, and  by introducing a phenomenological boundary at
$\sigma = \sigma_b$. We obtained stationary probability
distributions for a wide class of theories with two different
time parametrizations.

If nothing precludes sliding of the probability distribution
along the boundaries, one typically obtains a non-stationary
distribution.  However, the  local stationarity still exists:
The properties of domains with given values of scalar fields
$\phi$ and $\sigma$ do not depend on the time when these domains
were formed.

Whether the probability distribution $P_p$ is stationary or not,
in the process of its evolution it probes all   values of
the fields $\phi$ and $\sigma$ for which inflation and
self-reproduction of the Universe can take place. As a result,
after inflation the Universe becomes divided into many
exponentially large domains with different values of the
effective Planck mass $M_{\rm p}(\phi)$.

It would be natural to assume that the probability to live 
in a typical part of our
Universe  is proportional (though not equal, see below) 
to ${\cal P}_p(\phi_e)$. 
In particular, if our calculations would give us a delta-functional
distribution ${\cal P}_p(\phi_e)$, we would have a definite
prediction for $M_{\rm p}$ and thus for $G$. Our results show,
however,  that the distributions which we obtain for large
values of masses and coupling constants are rather smooth. If
one takes very small masses and coupling constants, which is
necessary to obtain small density perturbations
$\delta\rho/\rho$, the distributions become very sharply peaked
indeed, but still they are not delta-functional.

The probability to live in a given part of the Universe depends
not only on its volume (which is proportional to the distributions
$P_p$ and ${\cal P}_p$ if they are sufficiently narrow) 
but also on the conditions inside this
volume.  These conditions depend very strongly on the value of
$M_{\rm p}$. For example, it is well known that  a decrease of
the Planck mass $M_{\rm p}$ by less than an order of magnitude
from its present value in our part of the Universe would make
the lifetime of the Sun so small that no biological molecules
would appear on the Earth. An even bigger decrease of $M_{\rm
p}$ would lead to an extremely efficient nucleosynthesis and to
the absence of hydrogen in the Universe \cite{BT}. An increase
of $M_{\rm p}$ would slow the expansion of the Universe. In such
a Universe the departure from thermal equilibrium during the
process of baryogenesis would be small, this process would be
inefficient and the Universe now would be practically empty. On
the other hand, a decrease of $M_{\rm p}$ decreases the
reheating temperature after inflation, which may  also cause the
absence of baryons.  Finally, in the simplest model of inflation
with $V(\sigma) = {m^2\over2} \sigma^2$, density perturbations
produced during inflation are inversely proportional to $M_{\rm
p}$, see eq. (\ref{RPE}).  Therefore a change of $M_{\rm p}$
would lead to a profound modification of the properties of
galaxies.

This suggests that the knowledge of the distribution ${\cal
P}_p$ being complemented with anthropic considerations may help
us determine the most probable value of the gravitational
constant in the domains of  the Universe where life of our type
is possible. This is a very exciting possibility, resembling the
`big fix' paradigm of the baby universe theory \cite{Coleman},
but, just like the baby universe theory,  it involves many
speculations. One of the problems of such an approach is the
dependence of  ${\cal P}_p$ on the choice of time
parametrization. We will  briefly discuss this issue in the
Appendix. Whether this most ambitious part of our program will
be successful or not, it is certainly true that the theory of a
self-reproducing inflationary Brans--Dicke universe offers us
many interesting and unexpected possibilities.

For example,  in the standard inflationary cosmology the Planck
mass was fixed, and in order to obtain a desirable amplitude of
density perturbations ${\delta\rho\over\rho} \sim 5\cdot
10^{-5}$  in the theory ${m^2\over 2}\sigma^2$ one should
introduce into the theory a new mass scale, $m \sim 10^{13}$
GeV, which is six orders of magnitude smaller that the Planck
mass $M_{\rm p} \sim 10^{19}$ GeV. This cannot be considered as
a fine tuning; after all, the electron mass is 22 orders of
magnitude smaller than the Planck mass. Still it is very
interesting to see how the same issue looks in the context of
the inflationary Brans--Dicke theory.

In this theory $m$ is fixed but the Planck mass is not. It takes
all its possible  values in different parts of inflationary
universe.  Correspondingly, the amplitude of density
perturbations in different parts of the Universe takes all
possible values from $0$ to $O(1)$.  If by calculating ${\cal
P}_p$ and using anthropic considerations we are able to
explain why we live in a part of the Universe with
${\delta\rho\over\rho} \sim 5\cdot 10^{-5}$  (see Appendix),
then we  will simultaneously explain why ${\delta\rho\over\rho}$
is so small  and why $M_{\rm p}$ is so large in our part of the
Universe. But there is a chance that we  live in a part of the
universe with ${\delta\rho\over\rho} \sim 5\cdot 10^{-5}$
without any special reason, just as some people live in exotic
countries without even knowing that they are exotic.
Nevertheless, even in this case we will get something
interesting. We should simply look at equation (\ref{RPE}) in a
different way, writing it as follows:
\begin{equation}\label{RFE2}
M_{\rm p}(\phi)\sim 50\ m \cdot
\left({\delta\rho\over\rho}\right)^{-1}\ .
\end{equation}
This equation tells us that if  we  live in a part of the
Universe with ${\delta\rho\over\rho} \sim 5\cdot 10^{-5}$, then
in this part of the Universe the effective Planck mass
automatically happens to be one million times larger than  $m$,
which is the only mass scale we have in our simple theory. In
other words, we do not need to introduce into the theory two
different mass scales different from each other by a factor of
$10^6$. It is enough to have one mass scale; the rest of the job
will be accomplished by quantum fluctuations in the inflationary
universe. In this scenario both smallness of  the density
perturbations and greatness of the Planck mass appear  as  two
sides of the same purely environmental effect.

\vspace{1cm}

The authors express their gratitude to  A. Mezhlumian  and V.
Mukhanov for valuable discussions.  The work of A.L. was
supported in part by NSF grant PHY-8612280. The work of J.G.-B.
was supported  by  a Spanish Government MEC-FPI postdoctoral
grant.

\vfill
\newpage

\section*{\label{appen} Appendix. \ \ Towards determination of
the most probable value of the gravitational constant $G(\phi)$}

One of the most interesting problems of elementary particle
physics is to understand why the gravitational constant is so
small or equivalently the Planck mass $M_{\rm p}$ is so large.
In the context of our model this problem is formulated in a very
unusual way. Our Universe consists of many exponentially large
domains with different values of  $G = M_{\rm p}^{-2}(\phi)$.
Therefore instead of finding a unique value of $M_{\rm p}$ for
the whole Universe, one should study the distribution of all
possible values of $M_{\rm p}$ in our Universe.  This might give
us a possibility to understand   why the gravitational constant
is so small in the part of the Universe {\it where  we live}.

First of all, one can make a natural assumption that the number
of observers asking questions about the parts  of the Universe
with given values of fields $\phi$ and $\sigma$ is proportional
to the volume of these parts. This suggests that the answer to
our questions may be related to the investigation of the
distributions $P_p$ and ${\cal P}_p$. However, it is not clear
whether it is possible to justify this suggestion, especially if
one recalls that these distributions depend on the choice of the
time parametrization.

Note, that these probability distributions  have a very well
determined operational meaning when one uses them to predict the
distribution of a scalar field in the Universe at a specific
hypersurface   $t =const$ (or $\tau = const$) under given
initial conditions at $t=0$ (or at $\tau = 0$).  Now the main
question is whether it is possible to use the results of our
calculation of $P_p$ and ${\cal P}_p$ to get some information
about the most probable value of the gravitational constant in
those parts of the Universe where we can make observations and
ask questions about the gravitational constant.

One may argue that as far as we have a nonvanishing probability
to live in the domains with a given value of $M_{\rm p}(\phi)$,
and as far as the total volume of all such domains integrated
over all times is infinite (the integral exponentially diverges
at $t\to \infty$), there is no reason to compare these
infinities and study details of   behavior of $P_p$.  We live,
which means that we have picked up one of these domains, but all
attempts to go any further than that and to explain `our choice'
would not make any sense. This would be similar to attempts of a
man from Switzerland to understand why he was born there rather
than in China where the total number of people is much greater.

According to this point of view, one should use our results in a
very limited way. One should find those domains  where the
distribution ${\cal P}_p$    and the probability of existence of
life of our type do  not vanish. (The answer to this question
does not depend on the choice of time parametrization.) One
should then find out what is the value of the effective Planck
mass in each of these domains, and to relate it to other
properties of  space and matter in these domains.

In the Discussion we took this most conservative attitude
towards our results.  Even in this case inflationary cosmology
goes far beyond the standard Big Bang paradigm, which  assumes
that the gravitational constant should be the same in all parts
of our Universe.

This conservative attitude is quite legitimate. It is quite
possible indeed that  one should not ask why he or she was born
in this or that country and in this or that part of the
Universe.  However, this idea is suspiciously similar to the old
belief that it does not make any sense to question initial
conditions in the Universe: The Universe is big and flat for the
reason that it was born  big and flat; it contains more baryons
than antibaryons for the reason that it was created that
way...\, After the invention of the theory of baryogenesis and
of inflationary cosmology this way of avoiding complicated
problems does not look particularly attractive.

On the other hand, we are not well prepared to go beyond this
point.  In order to understand why do we live in a part of the
Universe with a given value of the gravitational constant, one
should learn first what is life, how it appears, is it correct
that the probability for life to appear is proportional to the
space available, etc.  Before we do it, there will be no
guarantee that we are on the right track. However, instead of
waiting until the theory of everything is constructed,  we may
use inflationary cosmology as a good playground where we can
test various hypotheses. This might help us to learn how to
formulate correct questions (and, if we are lucky, to get
correct answers) in the context of  the new cosmological
paradigm.

To illustrate some ambiguities involved in this approach, let us
consider  a problem formulated some time ago by Holger Nielsen
\cite{Nielsen}.  Assume that we live in a peak of probability to
be born at some particular time $t$. This hypothesis at the
first glance looks quite reasonable and innocuous.  In any case,
it   does not look obviously wrong   if we are looking to the
past. Indeed, the total population of the Earth now is much
larger than it was before, and it continues growing
exponentially.  Most of the people who have ever lived on the
Earth were born in  the  20th century. This gives us a total of
less  than 20 billion people.  If we assume that a typical person
should be born near the maximum of the probability distribution,
and if we assume that we are typical,  then we would not expect
much more that 20 billion people to be born from now on. But
this is possible only if very soon, within the next few decades,
the population of Earth will start rapidly decreasing
\cite{Nielsen}.

We definitely want this doomsday prediction to be wrong, but
what could be wrong about it? The point is that this prediction
is a consequence of the assumption that the total number of
people to be born is finite, and we live near the maximum of the
distribution.  However, if the total number of the people to be
born is infinite, then we live at some time $t$ not for the
reason that this time is near the maximum of the distribution,
but for the only reason that we must pick up some {\it finite}
time $t$ rather than the time $t = \infty$.  For example, in our
scenario  the total volume of the Universe grows exponentially
at all times. Consequently, the total number of domains of our
type, the total number of planets of the type of the Earth and
the total number of people populating these planets also grow
exponentially \cite{LLM}.  Thus, in this scenario the total
population of the Universe does not have any maximum and any
falloff at large $t$.

This example shows that one should be extremely careful with
probabilistic arguments to avoid many hidden ambiguities. For
this reason we removed this discussion from the main body of the
paper, to make sure that our speculations do not mix up with
reliable results.

One of the obvious problems with the distributions $P_p$ and
${\cal P}_p$ is the choice of the time parametrization. Indeed,
as we have seen, position of the peaks of these distributions
does depend on this choice. Of course, one may argue that
life as we know
it uses ordinary time $t$ related to periodic processes, rather
than the time $\tau$ which measures the logarithm of the
distance between galaxies. Does it suggest that the proper
distribution to study is $P_p(\sigma,\phi;t)$?  Let us make this
assumption for a moment and see what we will learn.

First of all, the peaks of the distributions $P_p$ for all
theories we studied correspond to the state with a maximal
possible value of the Hubble constant compatible with
the self-reproduction of inflationary domains. For all theories
$V(\sigma) \sim \sigma^n$ the peak of $P_p$ appears near the
upper boundary $\sigma_b$, and the peak of  ${\cal P}_p$ appears
near $\phi \sim \sigma_b \sqrt{2\over n}$, which gives
$M_{\rm p} \sim 2 \sigma_b \sqrt{\pi\over \omega}$. The boundary
$\sigma_b$ may correspond either to the place where $V(\sigma)$
changes its shape and becomes very steep, or to the place where
it becomes decreasing.

Now one should multiply the volume of the domains with given
$M_{\rm p}$ by the probability that life of our type can exist
in these domains.  The results look rather strange. Anthropic
arguments seem to allow variations of $M_{\rm p}$ at least by
one order of magnitude. Meanwhile typical distributions ${\cal
P}_p$ are extremely sharp if masses and coupling constants are
realistically small.  Multiplication of these distributions
typically gives a distribution which is sharply concentrated near
one of the boundaries of the anthropically allowed region. This
is very similar to the results obtained by Rubakov and
Shaposhnikov when they discussed   anthropic considerations in
the context of the baby universe theory  \cite{RubShap}.

The situation becomes even more complicated when one considers
models where the distributions $P_p$ and ${\cal P}_p$ are
non-stationary.   In such models the probability distribution
${\cal P}_p$ permanently moves towards infinitely large $\phi$.
This growth eventually outweighs all anthropic bounds and pushes
the distribution towards the region where life of our type could
exist only as an extremely unstable phenomenon. This would be
clearly incompatible with the results of   anthropic
considerations, which suggest that we live not far away from the
center of the anthropically allowed region of possible values of
$M_{\rm p}$.

One may try to interpret this conclusion as an argument against
Brans--Dicke cosmology,\footnote{There exist some arguments in
the context of the baby-universe theory that the Brans-Dicke
theory with an account taken of the wormhole effects is reduced
to the standard Einstein theory \cite{Worm}.} or at least
against those versions which lead to the runaway solutions for
$P_p$ and ${\cal P}_p$.  Another possibility is to use the
distribution in the time $\tau$, which is typically much less
sharp. Note that this time parametrization also has certain
advantages, see \cite{SB}. Still another possibility is that the
distributions $P_p$ and ${\cal P}_p$, being very useful for the
description of the global structure of inflationary universe
{\it at a given time}, cannot be used for the calculation of the
probability of life  appearing in the part of the Universe of
our type. Indeed, as we know from \cite{DeWitt}, the notion of
time (either $t$ or $\tau$) makes sense in the context of
quantum cosmology only {\it after} the appearance of observers.

It is excitingly interesting to participate in the investigation
of this problem. However, at the moment we do not even know
whether physics provides a wide enough framework to study
appearance of life, or some additional ingredients are needed
\cite{MyBook}. This is an extremely complicated and speculative
issue. We hope to return to its discussion in a separate
publication. In the meantime  we decided to restrict ourselves
to the conservative approach outlined in the Discussion.

\end{document}